\documentclass[structabstract]{aa}  
\usepackage{graphicx}
\usepackage{aalongtable,lscape}
\usepackage[figuresright]{rotating}
\usepackage{natbib}
\usepackage{txfonts}

\newcommand\nicev{\upsilon}
\begin{document}
   \title{Molecular gas in low-metallicity starburst galaxies:}

   \subtitle{Scaling relations and the CO-to-H$_2$ conversion factor\thanks{Based on observations carried out with the IRAM 30m Telescope. IRAM is supported by INSU/CNRS (France), MPG (Germany) and IGN (Spain).}}

   \author{R. Amor\'in 
          \inst{1},  
	  C. Mu\~noz-Tu\~n\'on\inst{2,3}, 
	  J.A.L. Aguerri\inst{2,3}, 
 \and
 P. Planesas\inst{4}
          }

 \offprints{R. Amor\'in \email{ricardo.amorin@oa-roma.inaf.it}}

   \institute{
      {INAF -- Osservatorio Astronomico di Roma, Via Frascati 33, I-00040  Monte Porzio Catone, Roma, Italy}  
       \and
{Instituto de Astrof\'isica de Canarias (IAC), V\'ia L\'actea S/N, E-38200 La Laguna, Tenerife, Spain}
\and 
{Departamento de Astrof\'isica, Universidad de La Laguna, E-38206 La Laguna, Tenerife, Spain}       
\and
      {Observatorio Astron\'omico Nacional (IGN), Alfonso XII 3, E-28014 Madrid, Spain} 
}         
   \date{Received ..., ...; accepted ..., ...}
 
  \abstract
   {Tracing the molecular gas-phase in low-mass star-forming galaxies becomes 
   extremely challenging due to significant UV photo-dissociation of CO molecules 
   in their low-dust, low-metallicity ISM environments.
   }
   {We aim at studying the molecular content and the star formation efficiency of 
   a representative sample of 21 Blue Compact Dwarf galaxies (BCDs) previously 
   characterized on the basis of their spectrophotometric properties. 
   }
   {We present CO (1-0) and (2-1) observations conducted at the IRAM-30m telescope. 
   These data are further supplemented with additional CO measurements and 
   multiwavelength ancillary data from the literature. 
   We explore correlations between the derived CO luminosities and several 
   galaxy-average properties. 
   } 
   {We detect CO emission in 7 out of 10 BCDs observed. For two galaxies these are 
   the first CO detections reported so far. We find the molecular content traced 
   by CO to be correlated with the stellar and H{\sc i} {masses}, star formation rate 
   (SFR) tracers, the projected size of the starburst and its gas-phase metallicity. 
   BCDs appear to be systematically offset from the Schmidt-Kennicutt (SK) law, showing 
   lower average gas surface densities for a given $\Sigma_{\rm SFR}$, and therefore 
   showing extremely low ($\la$\,0.1 Gyr) H$_2$ and H$_2+$H{\sc i} depletion timescales.  
  The departure from the SK law is smaller when considering H$_2+$H{\sc i} rather than H$_2$ 
  only, and is larger for BCDs with lower metallicity and higher specific SFR. 
  Thus, the molecular fraction ($\Sigma_{\rm H_2}/\Sigma_{\rm HI}$) and
  CO depletion timescale ($\Sigma_{\rm H_2}/\Sigma_{\rm SFR}$) of BCDs
  is found to be strongly correlated with metallicity. 
   {Using this and assuming that the empirical correlation found between the 
  specific SFR and galaxy-averaged H$_2$ depletion timescale of more metal-rich 
  galaxies {extends} to lower masses}, we derive a metallicity-dependent CO-to-H$_2$ 
  conversion factor  {$\alpha_{\rm CO, Z} \propto (Z/Z_{\odot})^{-y}$, 
  with $y=1.5(\pm 0.3)$ } in  {qualitative} agreement with previous 
  determinations, {dust-based measurements,} and  {recent} model
  predictions. 
 {Consequently, our results suggest} 
  that in vigorously star-forming dwarfs the fraction of H$_2$ traced by CO decreases 
  {by} a factor of about  {40} from $Z \sim Z_{\odot}$ to $Z \sim 0.1 Z_{\odot}$, 
  leading to a strong underestimation of the H$_2$ mass in metal-poor systems when a 
  Galactic $\alpha_{\rm CO, MW}$ is considered. 
{Adopting our metallicity-dependent conversion factor
  $\alpha_{\rm CO, Z}$ we find that departures from the SK law are
  partially resolved}.} 
{{Our results suggest that starbursting dwarfs have shorter
  depletion gas timescales and lower molecular fractions compared to
  normal late-type disc galaxies even accounting 
for the molecular gas not traced by CO emission in metal-poor
environments, raising additional constraints to model predictions}.
}
\titlerunning {CO in low metallicity starbursts}
\authorrunning{R. Amor\'in et al.}

   \maketitle
%

\section{Introduction}
 \label{s1}

Blue Compact Galaxies (BCDs) are low luminosity gas-rich systems with
an optical extent of a few kpc \citep{TM81}. They undergo intense bursts of star
formation, as evidenced by their blue colours and strong nebular
emission, with ongoing star formation rates (SFR) typically $\sim
0.1-10 M_{\odot}$yr$^{-1}$ \citep[e.g.,][]{GdP03}. 
BCDs span in a wide range of sub-solar metallicities 
0.02\,$\la$\,$Z/Z_{\odot}$\,$\la$\,0.5), including the most
metal-poor star-forming galaxies known in the local Universe
\citep[e.g.,][]{Terlevich91,KunthOstlin2000,Kniazev2004,Izotov2006,
P06,P08,Morales-Luis2011,Filho2013}.
These extreme properties lead originally to conjecture that they
are pristine galaxies experiencing at present the formation of their
first stellar population \citep{SS70}.  Subsequent work has shown,
however, that most ($>$95\%) BCDs are old systems ($>$\,5 Gyr) that underwent
previous starburst episodes \citep[][]{Gerola80,Davies88,SanchezAlmeida08}. 
This conclusion relies mostly upon the detection of a more extended, evolved 
low-surface brightness host galaxy 
\citep{P96a,P96b,C01a,C03,N03,N05,Caon2005,GdP05,Vaduvescu06,
HunterElmegreen06,amorin07,amorin09,micheva13}.  
Nonetheless, even though they are not pristine galaxies, nearby BCDs   
constitute ideal laboratories to study in great detail vigorous star 
formation and galaxy evolution under physical conditions that are 
comparable to those present in low-mass galaxies at higher redshift 
\citep[e.g.,][]{Amorin2015,Amorin2014b,Amorin2014c,Maseda2014,deBarros15}.

So far, our understanding of the main processes triggering and
regulating star formation activity in BCDs remains very limited. 
This is largely due to our poor knowledge of the physical mechanisms behind
starburst activity, as well as of the feedback induced by the massive 
star formation in the interstellar medium (ISM). 
The strong UV radiation and the mechanical energy from stellar winds 
and SNe are likely agents of feedback in starbursts like those 
present in BCDs, leading to the ejection of
metal-enriched gas into the galactic halo, limiting the star
formation and shaping the large-scale structure and kinematics of the
surrounding ISM \citep[e.g.,][]{MacLow99,Silich2001,GTT06,Recchi13}. 

Likewise, the triggering of the starburst in BCDs remains a 
puzzle: these low-mass, gas-rich galaxies generally lack density waves 
and only a small fraction of them are seen in strong tidal interactions or
merging with more massive companions \citep[e.g.,][]{CamposAguilar93,TellesTerlevich95,Pustilnik2001,Koulouridis2013}. 
The origin of their ongoing star formation activity has been predominantly 
associated with fainter interactions with low-mass companions \citep[e.g.,][]{Noeske2001,Brosch2004,Bekki2008}, cold-gas accretion 
\citep[e.g.,][]{SanchezAlmeida2014,SanchezAlmeida2015} and other, barely 
understood internal processes \citep[e.g.,][]{P96b,vanZee2001,HunterElmegreen2004}. 

Results from detailed surface photometry have shown that starbursting 
dwarfs are more compact, i.e., they have a higher stellar concentration,  
than more quiescent dwarf irregulars \citep[dIs; e.g.,][]{P96b,GdP05,amorin09}. 
Moreover, spatially resolved H{\sc i} studies  \citep{VZ98,vanZee2001,Ekta10,LopezSanchez2012,Lelli2014a,Lelli2014b}  
have shown both that the central gas surface density in BCDs can 
be a factor $\ga$2 higher and that their H{\sc i} gas kinematics  
is more disturbed than in dIs. 
Some of these studies suggest that inflows of gas onto a BCD 
might lead to a critical gas density and, in turn, to the ignition of star 
formation on a small ($\sim$1 kpc) spatial scale. 
In addition, it has been suggested that gravity-driven motions and torques 
induced by the formation and further interaction of large star-forming 
clumps may also lead to central gas accretion and angular momentum loss, 
feeding the  current starburst \citep{Elmegreen12}. 
Additional observational support to this picture has been 
recently provided by studies of chemical abundances and ionized gas 
kinematics in metal-poor starbursts 
\citep[e.g.,][]{amorin10b,amorin12,Zhao13,SanchezAlmeida2013,SanchezAlmeida2014} 

A complete understanding of the above scenarios must take into account 
the relation between the total gas content and the ongoing SFR.  
In a classical paper \cite{Schmidt59} suggested the existence of a power-law 
relation between gas and SFR volume densities in the Milky Way.
Later, this relation was calibrated in terms of surface densities 
by \cite{K98} for a large number of galaxies, including late-type disks 
and massive starburst galaxies. 
The tight correlation found between the disk-averaged SFR and the 
gas (atomic, molecular or the sum of both) surface density, is a power law 
usually referred to as the ``Schmidt-Kennicutt law'' (hereafter SK law) and is 
of the type $\Sigma_{\rm SFR}$\,$\propto$\,$\Sigma_{\rm gas}^N$, with an 
exponent $N$ typically ranging 1.3-1.5 \citep[e.g.,][]{Bigiel08,Kennicutt12}. 
Defining the star formation efficiency (SFE) as the formation rate of 
massive stars per unit of gas mass able to form stars, or 
SFE $\equiv \Sigma_{\rm SFR}/\Sigma_{\rm gas}$, the above power law implies 
that galaxies with higher gas surface density will be more efficient 
converting gas into stars. 

Recent studies have identified the cold molecular gas density to be 
the primary responsible for the observed star formation rates in 
star-forming dwarf galaxies \citep[e.g.,][]{Bigiel08,Schruba11}.  
This is in agreement with the fact that massive star clusters 
occur in giant molecular clouds. 
The accumulation of large amounts of cold molecular gas 
would therefore be a prerequisite for the ignition of a 
starburst, leading to assume that galaxies with 
vigorous star formation must contain large amounts of H$_2$. 
The knowledge of the H$_2$ mass, spatial distribution and physical 
conditions is therefore essential for understanding the star formation 
processes itself, its ISM chemistry and the galaxy evolution as a whole. 
In spite of their relevance and the extensive observational effort
conducted over the last decade 
\citep[see e.g.,][]{Leroy05,Leroy11,Bigiel08,Bolatto11,Schruba12,Boselli2014,Cormier2014}, 
these remain key open questions for star-forming dwarf galaxies. 
\begin{table*}[t]
\caption{CO observations}
\label{T1}
\centering
\begin{tabular}{l l c c c c c c c}
\noalign{\smallskip}
\hline\hline
\noalign{\smallskip}
Galaxy & Other names & RA & DEC & D & $t_{\rm int}$ & $\theta_{\rm b}$ & $S/N$  \\
(1) & (2) & (3) & (4) & (5) & (6) & (7) & (8) \\ 
\noalign{\smallskip}
\hline
\noalign{\smallskip}
Haro\,1 & NGC\,2415 & 07 36 56.4 & 35 14 31.0 & 52.1 & 26  &5.46 & 6  \\[3pt]
Mrk\,33 & Haro\,2 & 10 32 31.9 & 54 24 03.5 & 22.3 & 52  &2.32 &  6   \\[3pt]
Mrk\,35 & Haro\,3 & 10 45 22.4 & 55 57 37.5 & 15.6& 104 &1.62 &   4  \\[3pt]
Mrk\,36 & Haro\,4 & 11 04 58.5 & 29 08 22.1 & 10.4 & 106 &1.08 &  $<$3 \\[3pt] 
Mrk\,297 & NGC\,6052 & 16 05 12.9 & 20 32 32.5 & 65.1& 27  &6.84 & $>$10 \\[3pt] 
Mrk\,314 & NGC\,7468 & 23 02 59.3 & 16 36 18.9 & 29.0 & 106 &3.02 & $<$3  \\[3pt]
Mrk\,324 & UGCA\,439 & 23 26 32.8 & 18 16 00.0 & 22.4 & 80  &2.33 & $<$3   \\[3pt]
Mrk\,401 & NGC\,2893 & 09 30 17.0 & 29 32 24.0 & 24.1 & 27  &2.50 & $>$10 \\[3pt]  
III\,Zw\,102& NGC\,7625 & 23 20 30.1& 17 13 32.0& 22.7 &  27 &2.36 & $>$10 \\[3pt]
III\,Zw\,107& PGC\,071605 & 23 30 09.8& 25 32 00.2& 78.1 & 168 &8.23 & {$<$3}   \\[3pt]
\noalign{\smallskip}								
\hline		
\hline										
\end{tabular}
\begin{list}{}{}
\item Notes.$-$ Columns: (3) and (4) equatorial coordinates, right
  ascension and declination ($J2000$), of the central pointing; (5)
  distance in Mpc; (6) total integration time in minutes; (7) CO
  ($J=1\rightarrow0$) beam-size (HPBW) in kpc; 
  (8) signal-to-noise ratio of the peak CO ($J=1\rightarrow0$) line.
\end{list}
\end{table*}

Due to the lack of a permanent dipole, H$_2$ emission arises only from 
hot or warm gas. For this reason, indirect methods are used to
estimate the mass of the \textit{cold} H$_2$ phase of the ISM. 
The most widely applied tracer is the CO rotational line emission.  
However, the cold phase of H$_2$, as traced by the $^{12}$CO molecule 
at millimetre wavelengths, seems to be mostly elusive in galaxies with 
metallicity below $\sim$\,20\% solar \citep[e.g.,][]{Elmegreen2013,Rubio2015}.  
In particular, BCDs are not only metal-poor but also strongly star-forming,  
a combination of properties that seems to disfavour high CO detection rates 
\citep[$\la$25\%,][]{Israel05, Leroy05}.  
This problem becomes particularly severe in the lowest-metallicity 
BCDs ($Z$\,$\la$\,0.1\,$Z_{\odot}$) for which only upper limits 
in CO luminosity exists and for which the CO-to-H$_2$ conversion 
factor $X_{\rm CO}$ seems to be extremely uncertain, e.g., in
the most metal-poor BCDs I Zw\,18 and SBS\,0335-052
($Z$\,$\sim$\,0.02\,$Z_{\odot}$), 
where even very deep searches for CO have proved fruitless \citep{Leroy07,Hunt2014}. 

 {Low metallicity environments imply lower C and O abundances and low dust-to-gas 
ratios \citep{Draine2007}. This affects the relative CO to H$_2$ abundances since dust is 
the site of H$_2$ formation and also provides much of the far-UV shielding 
necessary to prevent CO --that is not strongly self-shielding-- from 
photodissociating \citep{Bolatto13}.}
Most recent model predictions point to a CO deficiency in low-metallicity  
star-forming galaxies as due to a decrease of dust-shielding, which 
leads to strong photo-dissociation of CO by the intense UV radiation 
fields generated in the star-forming regions \citep[e.g.,][]{Wolfire10,Gnedin&Kravtsov10,Glover11}. 
Most of these models predict that, while the H$_2$ clouds survive -- via 
self-shielding or dust shielding -- to extreme, metal-poor ISM conditions, 
CO molecules are increasingly destroyed and therefore the conversion
factor $X_{\rm CO}$ depends strongly on the dust content and metallicity 
\citep[e.g.,][]{P&P09,Wolfire10,Glover11,Krumholz11,Dib11,Narayanan12}.  

In this paper we discuss the CO gas content and its relation with 
the main galaxy-averaged properties of a large sample of BCDs 
previously studied in a series of papers 
\citep[][and references therein]{amorin09}.  
To this aim, we present a small survey of the lowest
rotational transitions of $^{12}$CO for a sample of ten BCDs. 
These new data and additional CO measurements obtained from the literature 
for another eleven BCDs, are combined with a large multiwavelength ancillary 
data including metallicities, stellar and H{\sc i} masses and several SFR tracers. 
These are used to explore scaling relations between the CO luminosity  
and star formation in BCDs and, in particular, to study the apparent  
dependence of the molecular and total (H{\sc i}$+$H2) gas depletion 
timescales with metallicity. 
In this analysis, we are able to derive a tight  {power-law relation for}  
$X_{\rm CO}$ and metallicity, valid for vigorously star-forming dwarfs over 
nearly one order of magnitude in metallicity. 

We organize this paper as follows. In Section~2 and 3 we describe the sample 
of galaxies, the CO observations, and the ancillary datasets. 
In Section~4 we present our results, including the derivation of the
CO luminosities, masses and surface densities, and we study different
correlations with other galaxy-averaged properties, such as
metallicity, sizes 
and stellar and gas masses. 
 {In Section~5 we study the position of the BCDs in the star formation
   laws, while in Section~6 we discuss these results in terms of the
   star formation efficiency, specific star formation rate and
   metallicity. Later, in Section~7, we use these results to derive a
   metallicity dependent CO-to-H$_2$ conversion factor and derive total
   molecular masses. 
We then  revisit different scaling relations which are compared with
previous observational studies and discussed in terms of recent
models. 
Finally, in Section~8 we summarize our conclusions.}
Throughout this paper we adopt $H_0=$ 75 km\,s$^{-1}$\,Mpc$^{-1}$, 
$\Omega_0=$ 0.3, $\Omega_\lambda =$ 0.7, and solar metallicity 
12$+\log($O/H$)=$\,8.7 \citep{AllendePrieto2001}. 

\section{Sample of galaxies and CO observations}
\label{s2}

The sample of galaxies consists of {21} out of 28 BCDs presented and
studied by \citet{C01a,C01b}, which were characterized later in the
basis of their spectrophotometric properties in subsequent works 
\citep{C03,Caon2005,C07,amorin07,amorin08,amorin09,amorin10a}. 
As discussed in these previous studies, this sample was originally
selected as being representative of the BCD class. 
Thus, it covers the entire range of morphological types, luminosities,
colors, gas content, star formation activity and metallicity
properties seen in most BCD samples of the literature.  

\begin{figure*}[t]
\centering
\includegraphics[width=18cm,angle=0]{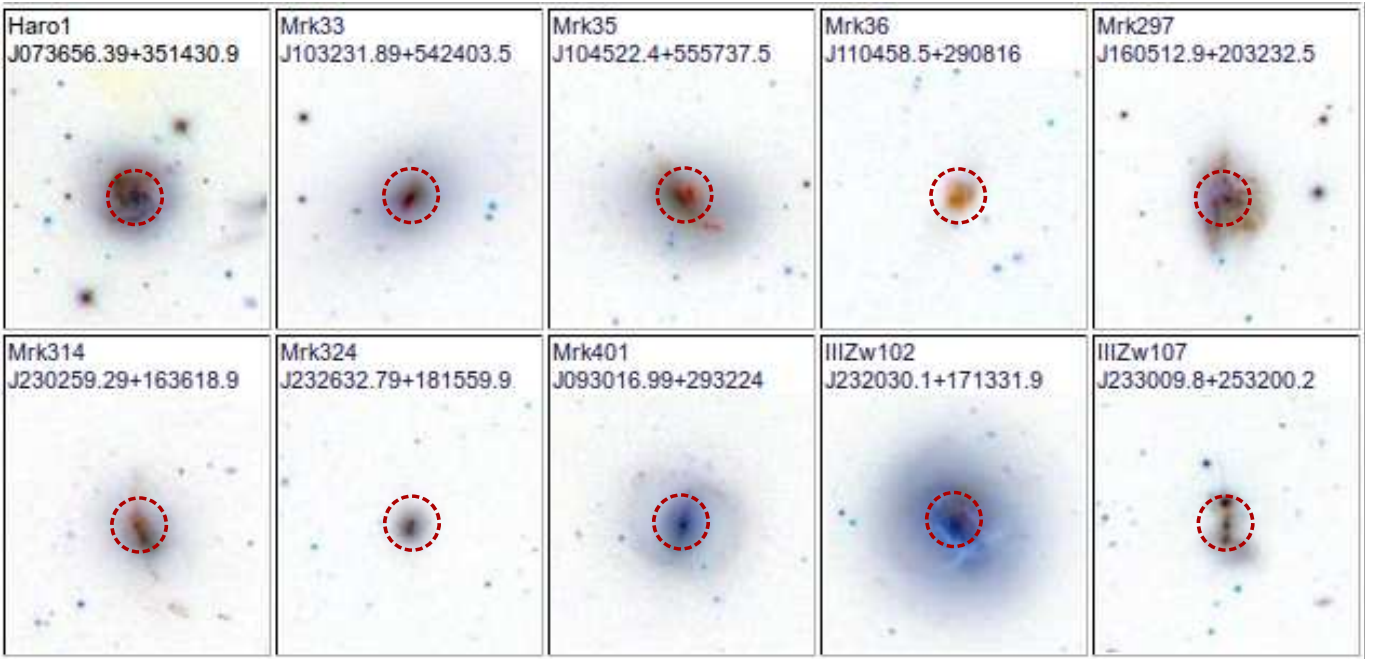}
\protect\caption[Sample of BCDs]{\footnotesize{Inverse \textit{ugriz} color-composite SDSS-DR10 thumbnails  of the observed galaxies. Each postage-stamp has 2$'$ on a side and a standard NE orientation, while the central circle indicates the center and size of the 22$''$ CO ($J=1\rightarrow0$) beam.}}
\label{fig1}
\end{figure*}

For the present study, we have conducted CO observations for a subset 
of 10 BCDs (hereafter referred to as {\it subsample~I}). 
{\it Sloan Digital Sky Survey} (SDSS) false-color thumbnails for 
these galaxies are shown  Fig.~\ref{fig1}. 
In Table~\ref{T1} we summarize the basic observational information for the 
observed galaxies.  
For the remaining {11} BCDs in the sample (hereafter referred to as {\it subsample~II}) 
we have compiled CO measurements from several sources in the literature, as described in  Section~\ref{s3}. 

The observed BCDs have been selected to favour their CO detectability 
in one or few pointings. They are small in size (optical diameter $<$2$'$), 
with declinations $\delta >$15\mbox{$^{\circ}$} and show blue colours and 
strong \mbox{H$\alpha$} emission in their centres, indicative of the presence 
of intense star formation activity. 
%
%

The observations of the $^{12}$CO $J$\,$=$\,1$\rightarrow$0 [$\nu$\,115 GHz] and
$J$\,$=$\,2$\rightarrow$1 [$\nu$\,230 GHz] rotational transitions were obtained
during April 25-28, 2006 at the IRAM 30~m telescope (Pico Veleta, Spain). 
The IRAM telescope provides beam sizes (HPBW\footnote{HPBW\,[arcsec] =
  2460/$\nu$\,[GHz], where $\nu$ is the observed frequency}) of about
22$''$ and 11$''$ at 115 GHz and 230 GHz, respectively, which 
{allowed} us to measure {the} CO emission on the sample galaxies over 
physical sizes between $\sim$1 and 8 kpc. 
In the most compact objects, where the star forming regions are 
located within the host galaxy effective radius 
\citep[\mbox{$r_{\rm e}$}$\sim$\,1-2 kpc, ][]{amorin09}, a single central 
pointing was enough to cover the whole CO emission (see Fig.~\ref{fig1}). 
On the other hand, in more extended BCDs the observed CO emission 
corresponds only to the most luminous regions of the starbursts, so it 
may be considered as a lower limit to the total CO emission. 
For all the observed galaxies our CO observations cover at least 
$\sim$\,50\% of the projected size of starburst ($r_{\rm SB}$, see Section~3), while for three objects this fraction, 100$\times$\,$\theta_{\rm b}/2r_{\rm SB}$, is $\ga$\,100\%.    

A wobbler switching mode was used at a frequency of 0.5~Hz 
with a beam throw up to 240$''$ in azimuth. Two independent
SIS receivers at each frequency (A100, B100 at 3mm and A230, B230 at
1.3mm) were used to observe both polarizations of the CO
$J$\,$=$\,1$\rightarrow$0 and $J$\,$=$\,2$\rightarrow$1 transitions 
simultaneously.
The receivers were tuned in single side band, using only the lower
band (LSB), which improves the calibration procedure and avoids
contamination by other spectral lines in the image band, especially
in observations of the calibration sources. The SIS receivers were
connected to two filter-banks with resolutions of 1~MHz and 4~MHz at
115~GHz and 230~GHz, respectively, which yields velocity resolutions,  
$\delta\nicev$, of 2.6 km\,s$^{-1}$ and 5.2 km\,s$^{-1}$, respectively. 
These backends have 512 channels each, covering a total instantaneous 
velocity range of 1300 and 2600 km\,s$^{-1}$.
Pointing and focus were checked before each integration on continuum
sources and planets. The pointing accuracy was better than
2.5$''$ on average. Calibrations were done using the chopper wheel method, by
observing strong radio sources (Orion~A, CW Leo, and W51d). System
temperatures {at} both frequencies varied between $\sim$\,290-470~K,
the higher values due to bad weather conditions. Thus, we obtained
a range for the antenna temperature sensitivity of $\delta T_{\rm rms}$\,$\sim$\,5-15\,mK.
\begin{figure*}[t]
\centering
\begin{tabular}{l l l l}
\includegraphics[width=4.35cm,angle=0]{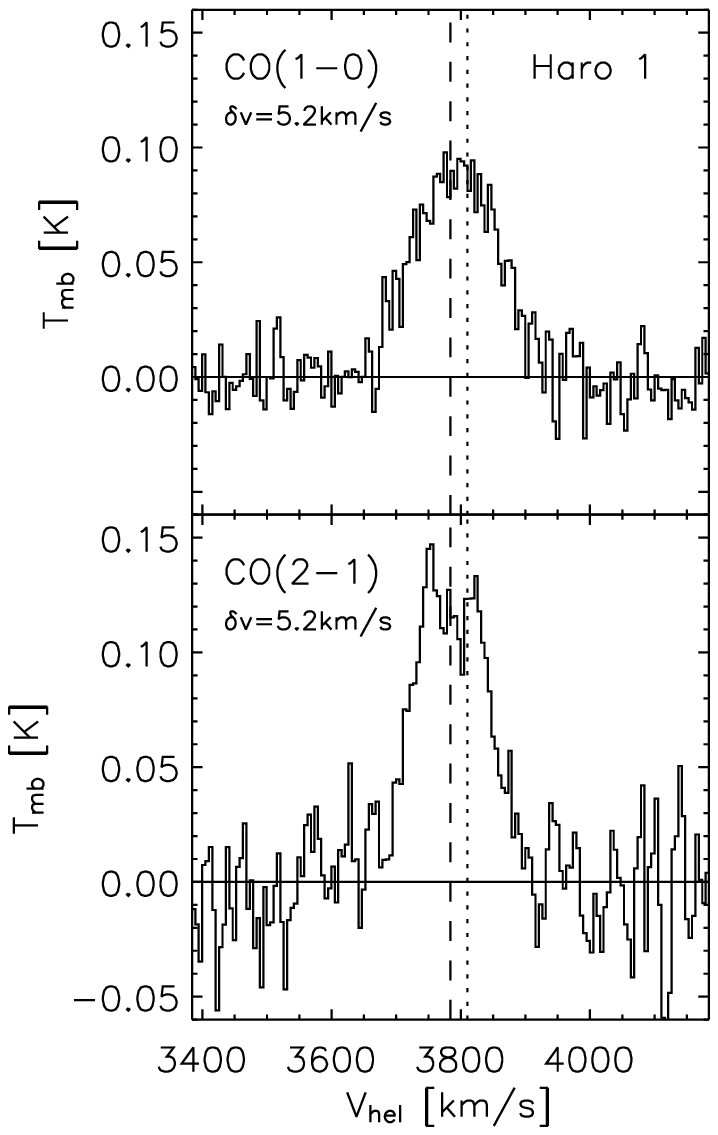}&
\includegraphics[width=4.35cm,angle=0]{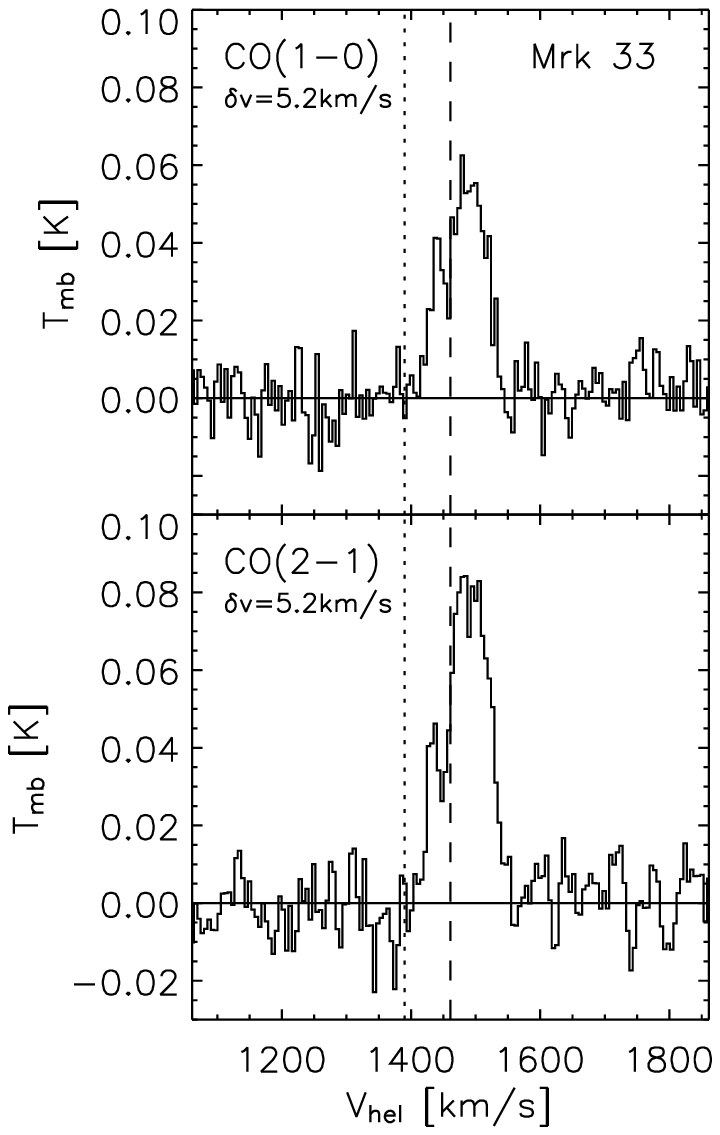}&
\includegraphics[width=4.28cm,angle=0]{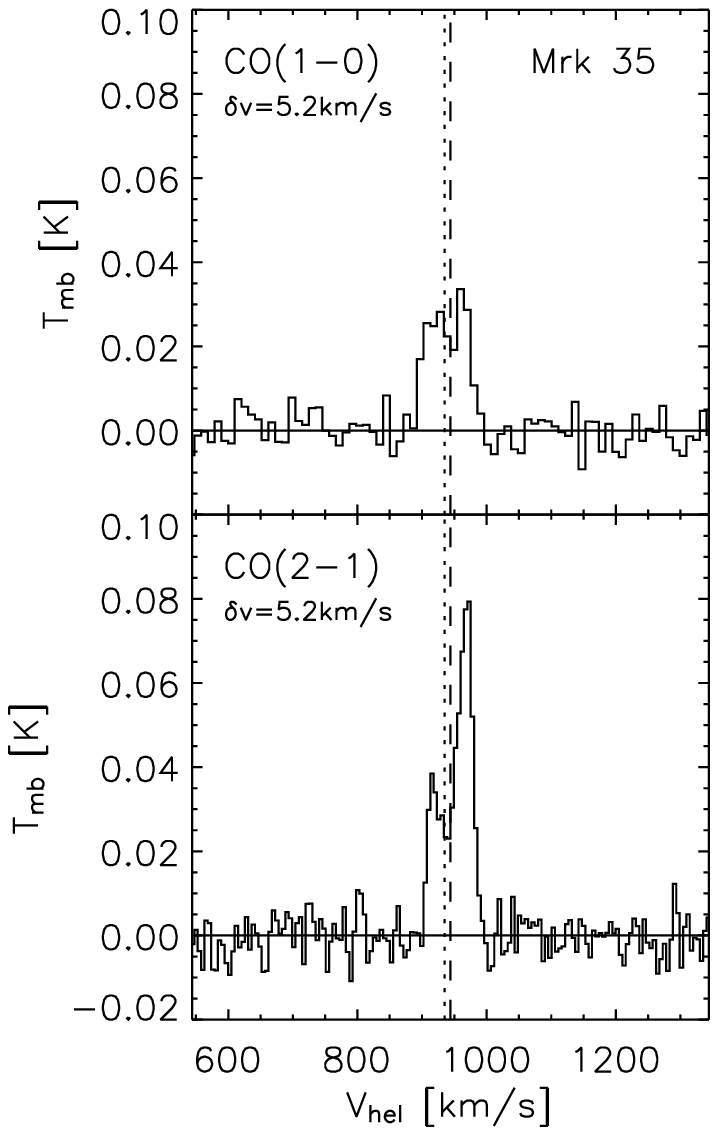}\\[8pt]
\includegraphics[width=4.30cm,angle=0]{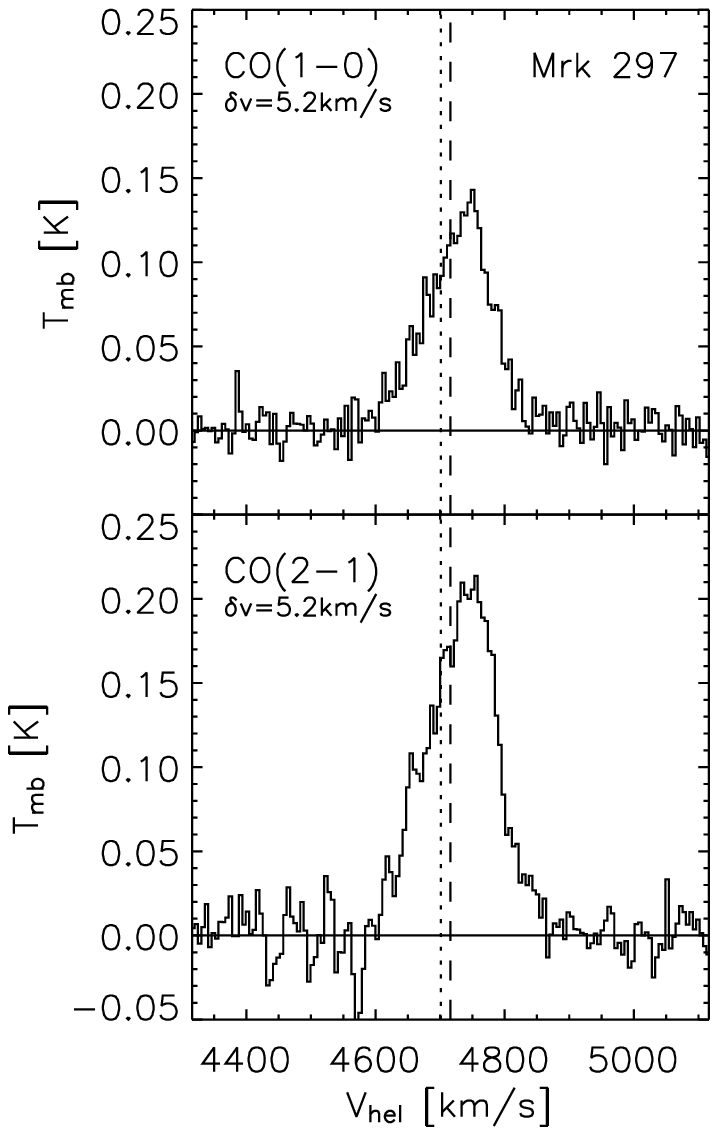}& 
\includegraphics[width=4.00cm,angle=0]{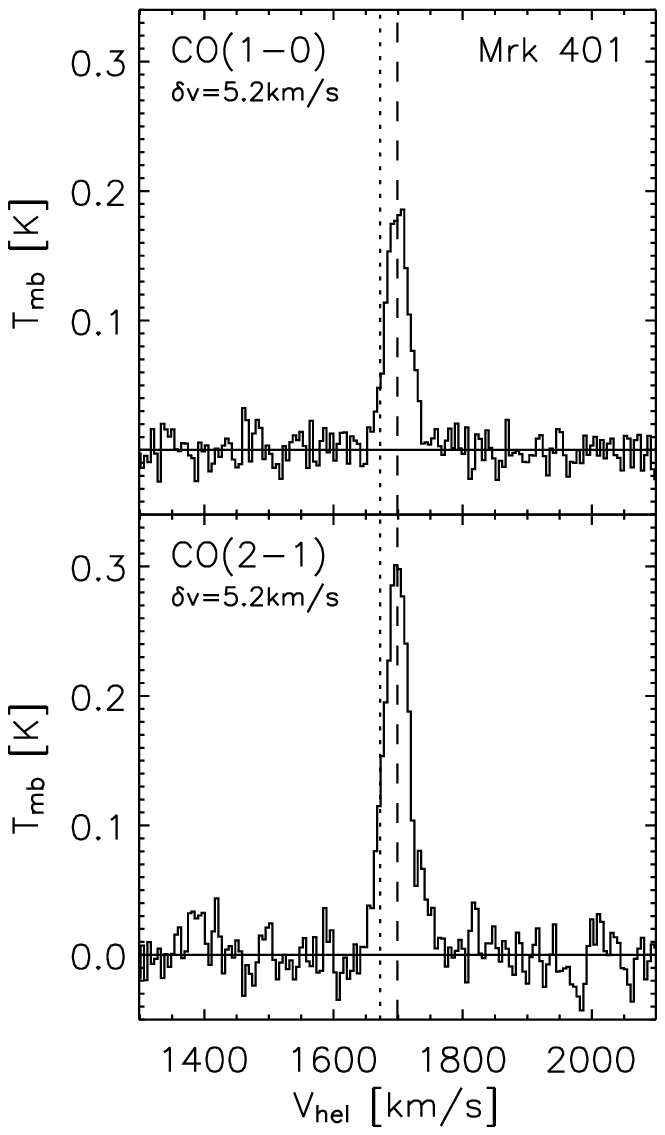}&
\includegraphics[width=4.15cm,angle=0]{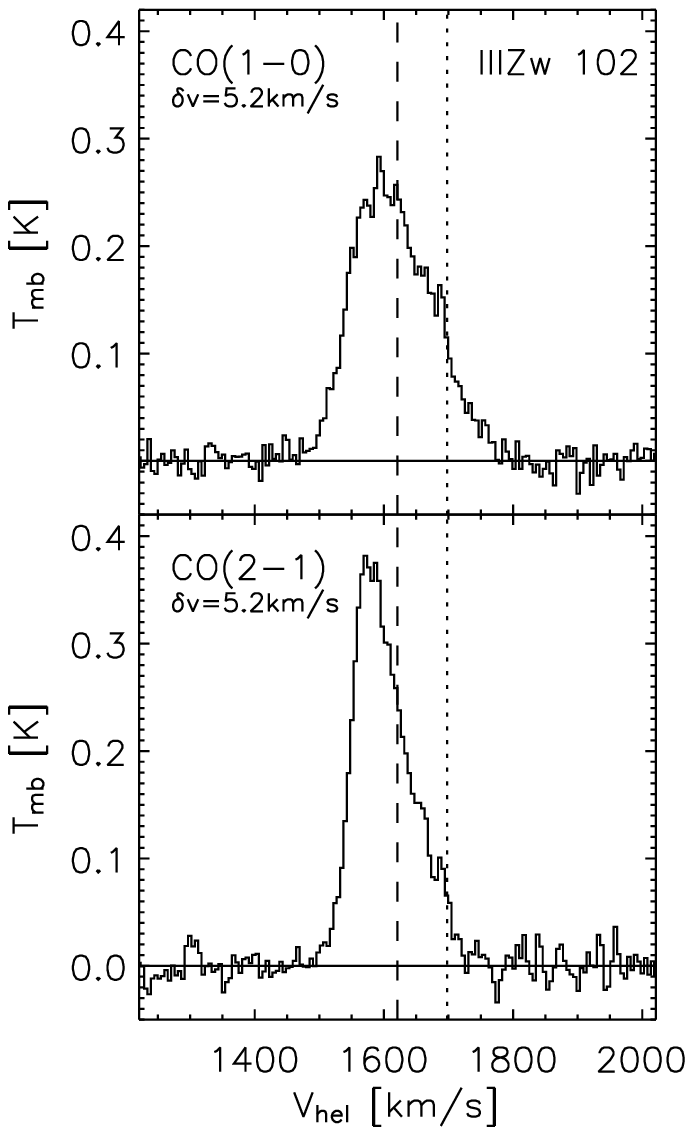}\\
\end{tabular}
\protect\caption[CO detections)]{\footnotesize{CO detections:  
CO $J=1 \rightarrow 0$ and $J=2 \rightarrow 1$ spectra are
shown for the target galaxies. Horizontal lines indicate the
adopted baseline, while vertical lines mark the systemic (heliocentric) 
velocities as measured from the optical (dotted line) and in 21~cm 
(dashed lines). The velocity resolution $\delta\nicev$, CO transition, and 
the name of the galaxy are also labelled in each plot.}}
\label{fig2a}
\end{figure*}

The CO spectra was reduced using CLASS\footnote{The CLASS software
  package is part of the GILDAS software, and can be downloaded from
  the IRAM website: http://www.iram.fr/IRAMFR/GILDAS/}
\citep{buisson}. In order to increase the signal-to-noise ratio (S/N)
averaged spectra for each pointing was obtained weighting multiple
scans by a factor of $t/T_{\rm sys}^2$, where $t$ is the integration
time and $T_{\rm sys}$ is the system temperature. In all cases,
baselines of zeroth order were subtracted. The spectra was
then scaled to a main-beam brightness temperature, $T_{\rm mb} =
\frac{F_{\rm eff}}{B_{\rm eff}} {T_{A}}^{*}$, where $F_{\rm eff}$ and
$B_{\rm eff}$ were the forward and beam efficiencies appropriate to
the epoch of observations. These values were 0.95 and 0.75 at 115 GHz,
and 0.91 and 0.52 at 230 GHz. Finally, all spectra were smoothed to a
resolution $\delta\nicev$ of 5~km s$^{-1}$ or 10~km s$^{-1}$, depending 
on the S/N. 
\begin{figure*}[ht!]
\centering
\begin{tabular}{l l l l}
\includegraphics[width=4.32cm,angle=0]{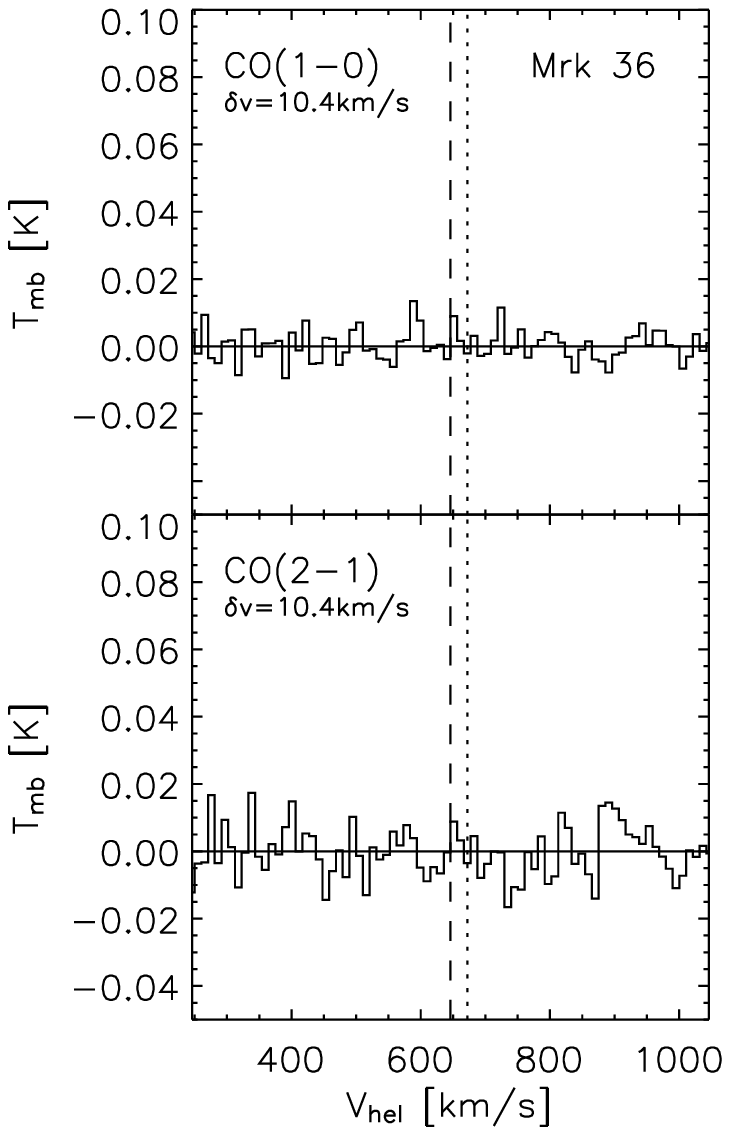}&
\includegraphics[width=4.25cm,angle=0]{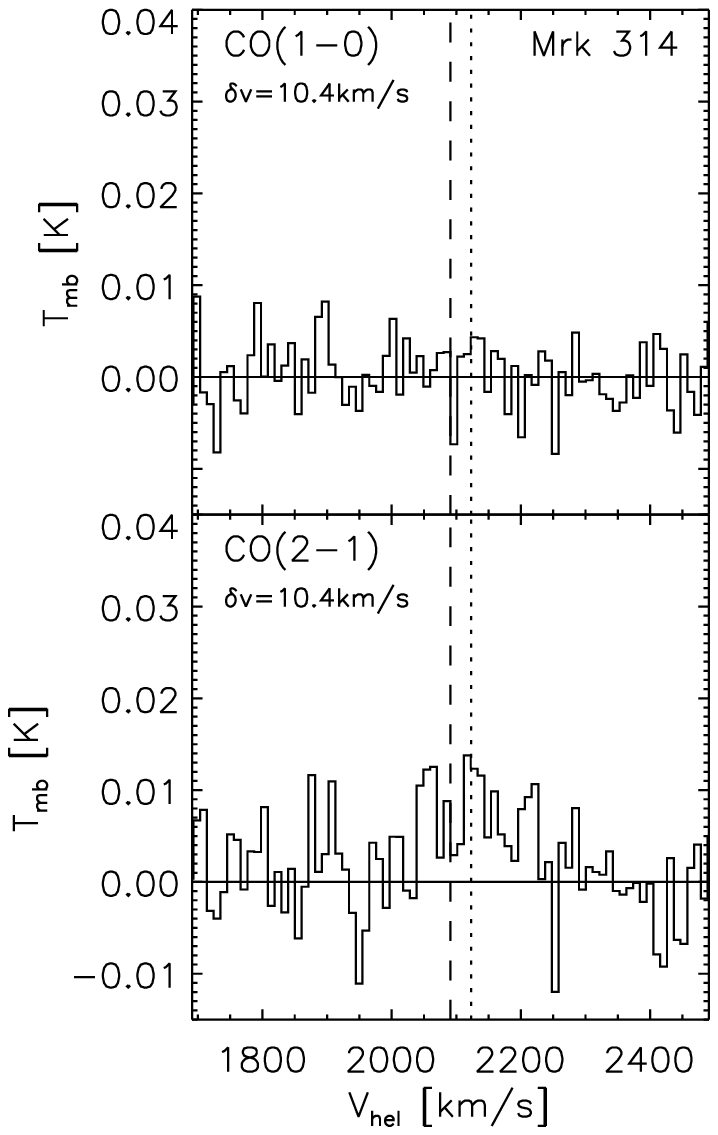}&
\includegraphics[width=4.50cm,angle=0]{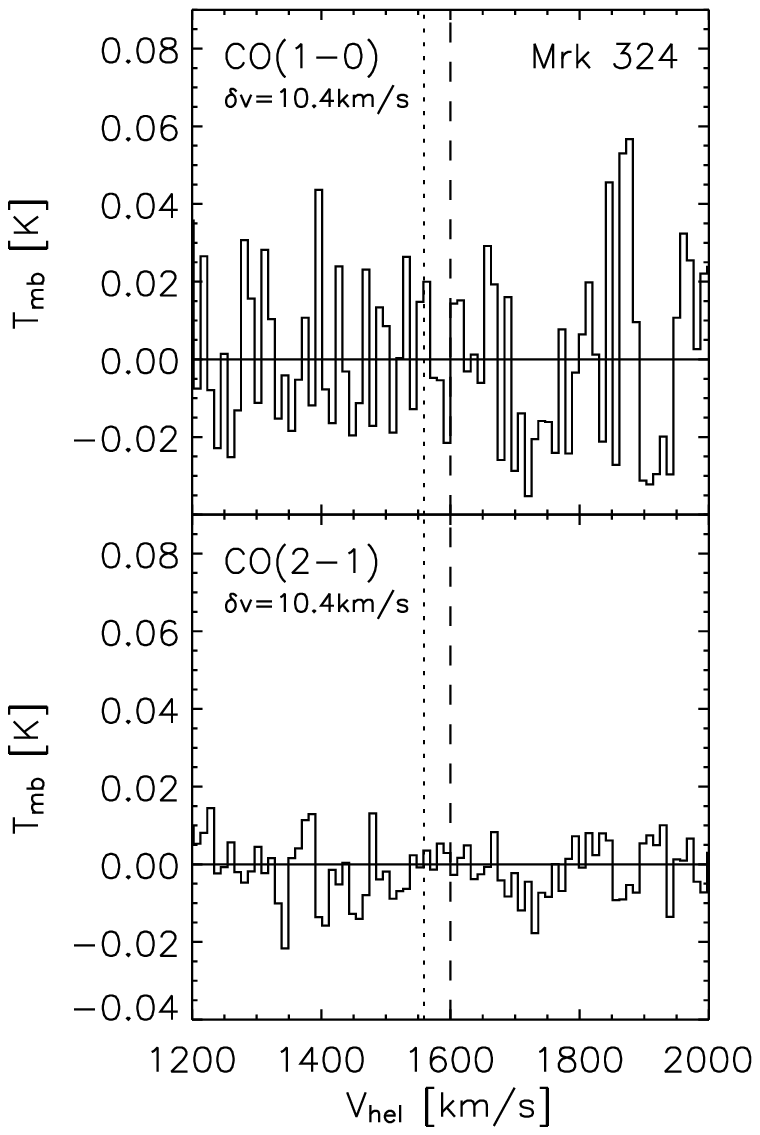}&
\includegraphics[width=4.30cm,angle=0]{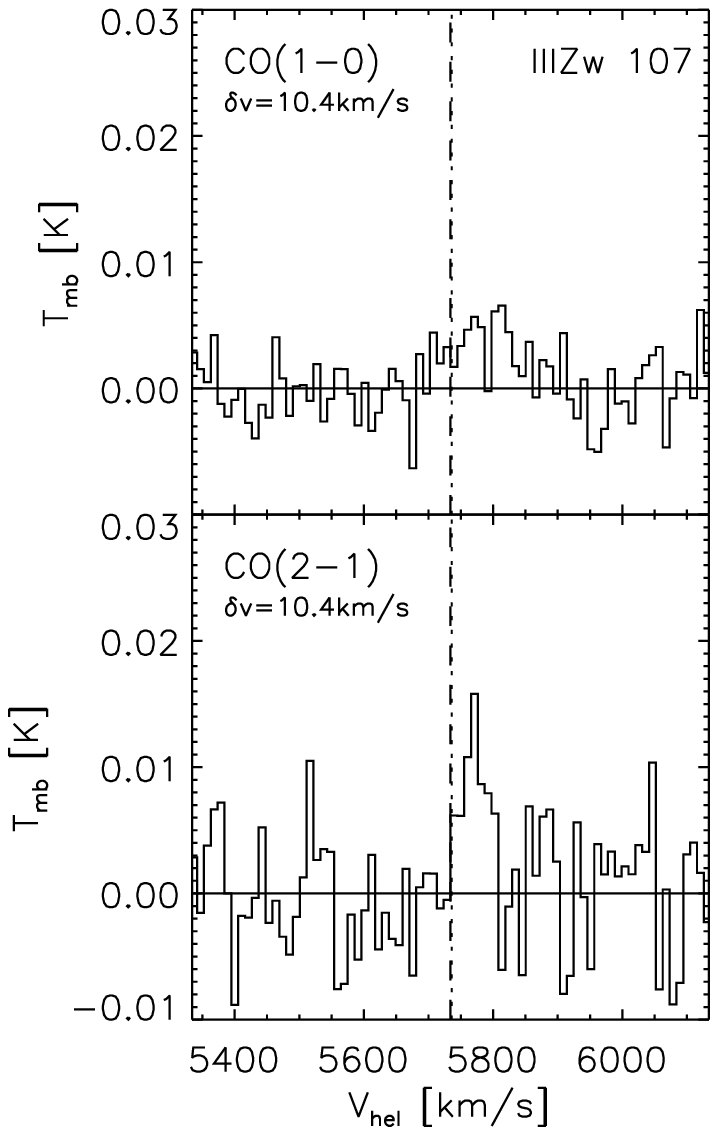}\\
\end{tabular}
\protect\caption[CO non-detections)]{\footnotesize{CO {tentative detections}  and non-detections: symbols, axis and lines are as in Fig~\ref{fig2a}.}}
\label{fig2b}
\end{figure*}

\section{Ancillary datasets}
\label{s3}

\begin{table*}[ht!]
\caption{{Multiwavelength properties of the observed galaxies}}
\label{T1.A1}
\centering
\begin{tabular}{l c c c c c c c c}
\noalign{\smallskip}
\hline\hline
\noalign{\smallskip}
Galaxy &  $r_{25}$ & $r_{\rm SB}$ &		
$\sin i$ & M$_{\rm HI}$ & $\log \Sigma_{\rm HI}$ &  
M$_{\star}$ & $\log \Sigma_{\rm SFR}$  & 12$+\log$(O/H) \\[3pt]
& kpc & kpc &  & 10$^9$\,M$_{\odot}$ &  M$_{\odot}$\,pc$^{-2}$ & 10$^9$\,M$_{\odot}$ & 
$M_{\odot}$\,yr$^{-1}$\,kpc$^ {-2}$ & dex \\ [3pt]
  & (2) & (3) & (4) & (5) & (6) & (7) & (8) & (9) \\
\noalign{\smallskip}
\hline
\noalign{\smallskip}
Haro\,1   & 6.98& 3.33& 0.19& 6.48 & 1.32 & 6.92& -0.52 &  8.53$^{a}$ \\[3pt]
Mrk\,33   & 3.25& 1.72& 0.72& 0.54 & 0.75 & 7.41& -1.05 &  8.35$^{a}$ \\[3pt]
Mrk\,35   & 3.07& 1.57& 0.71& 0.68 & 0.91 & 2.34& -1.10 &  8.30$^{b,c}$ \\[3pt]
Mrk\,36   & 0.45& 0.51& 0.72& 0.03 & 1.25 & 0.11& -1.10 &  7.85$^{a}$ \\[3pt]
Mrk\,297  & 8.80& 6.31& 0.64& 11.84& 1.28 & 5.62& -0.41 &  8.54$^{c}$ \\[3pt]
Mrk\,314  & 3.83& 1.51& 0.82& 2.50 & 1.19 & 0.38& -1.40 &  8.11$^{c}$ \\[3pt]
Mrk\,324  & 0.82& 0.51& 0.61& 0.23 & 1.62 & 0.91& -1.15 &  8.18$^{e}$ \\[3pt]
Mrk\,401  & 3.92& 1.69& 0.65& 0.82 & 0.81 & 1.12& -1.30 &  8.86$^{\zeta}$  \\[3pt]
III\,Zw\,102& 5.22& 2.45& 0.24& 1.98 & 1.17 &15.85& -0.64 &  8.74$^{c}$ \\[3pt]
III\,Zw\,107& 6.82& 3.97& 0.78& 6.69 & 1.16 & 2.00& -1.22 &  8.23$^{d}$ \\[3pt]
\noalign{\smallskip}							     
\hline							      			
\hline				      					
\end{tabular}									
\begin{list}{}{}
\item Columns: (2) radius at the $\mu_{\rm B}=$\,25 mag arcsec$^{-2}$
  isophote; (3) radius of the star-forming region from 2D surface photometry \citep{amorin09};   
  (4) inclination; (5) H{\sc i} gas mass; 
  (6) logarithm of H{\sc i} gas surface density; (7) stellar mass; 
  (8) logarithm of star formation rate density; (9) gas-phase metallicity; 
    {references} for metallicity: $(a)$ \citet{Shi06}, $(b)$ \citet{C07}, 
  $(c)$ \citet{Bego08}, $(d)$ \citet{Angel10}, $(e)$ \citet{Zhao10}, 
  $(f)$ \citet{PMD03}, $(g)$ \citet{GdP03}, ($\zeta$) this work.
\end{list}
\end{table*}			
%
%
\begin{table*}[ht!]
\caption{Multiwavelength properties of galaxies from the literature}
\label{T2.A1}
\centering
\begin{tabular}{l c c c c c c c c c}
\noalign{\smallskip}
\hline\hline
\noalign{\smallskip}
Galaxy & $r_{25}$ & $r_{\rm SB}$ & 
$\sin i$ & M$_{\rm HI}$ & $\log \Sigma_{\rm HI}$ &  M$_{\star}$ 
& $\log \Sigma_{\rm SFR}$ & 12$+\log$(O/H) & $\log \Sigma_{\rm H_2}$ \\[3pt]
 & kpc & kpc &  & 10$^9$\,M$_{\odot}$ & M$_{\odot}$\,pc$^{-2}$ & 10$^9$\,M$_{\odot}$ 
 & M$_{\odot}$\,yr$^{-1}$\,kpc$^{-2}$ & dex & M$_{\odot}$\,pc$^{-2}$ \\ [3pt]
  & (2) & (3) & (4) & (5) & (6) & (7) & (8) & (9) & (10) \\
\noalign{\smallskip}
\hline
\noalign{\smallskip}
Haro\,15  & 4.03& 3.55& 0.78& 5.5 & 1.06& 9.92& -1.28& 8.33$^a$ & $<$0.62$^h$ \\[3pt]
Mrk\,5    & 3.04& 0.74& 0.74& 0.14& 0.70& 0.07& -1.72& 8.10$^a$ & $<$0.47$^i$ \\[3pt]
Mrk\,86   & 10.3& 1.37& 0.49& 0.05& 0.09& 1.75& -1.70& 8.53$^g$ &    0.54$^k$ \\[3pt]
Mrk\,370  & 7.01& 1.01& 0.70& 0.29& 0.74& 0.83& -1.66& 8.51$^c$ &    0.44$^i$ \\[3pt]
Mrk\,1089 & 6.04& 3.82& 0.79& 7.46& 1.51& 3.38& -1.37& 8.22$^d$ &    0.61$^l$ \\[3pt]
Mrk\,1090 & 2.69& 1.66& 0.00& 1.94& 1.33& 1.05& -0.92& 8.15$^d$ & $<$0.25$^m$ \\[3pt]
UM\,462   & 3.47& 0.92& 0.72& 0.22& 1.15& 7.95& -0.80& 7.99$^f$ &    0.22$^j$ \\[3pt]
I\,Zw\,123& 1.39& 0.43& 0.47& 0.06& 1.02& 0.12& -1.12& 7.86$^a$ & $<$0.55$^i$ \\[3pt]
II\,Zw\,40& 1.40& 0.92& 0.79& 0.45& 1.35& 8.09& -0.69& 8.07$^a$ &    0.62$^n$ \\[3pt]
II\,Zw\,70& 3.61& 1.08& 0.97& 0.43& 0.66& 0.10& -1.70& 7.69$^a$ & $<$0.27$^m$ \\[3pt]
II\,Zw\,71& 4.05& 1.46& 0.92& 0.95& 1.03& 0.65& -1.77& 8.24$^a$ &    0.32$^p$ \\[3pt]
\noalign{\smallskip}							     
\hline							      			
\hline				      					
\end{tabular}									
\begin{list}{}{}
\item Columns: (2) radius at the $\mu_{\rm B}=$\,25 mag arcsec$^{-2}$
  isophote; (3) radius of the star-forming region from 2D surface photometry \citep{amorin09};  
  (4) inclination; (5) H{\sc i} gas mass; (6) logarithm of H{\sc i} gas surface density; (7) stellar mass; (8) logarithm of star formation rate density; (9) gas-phase metallicity; (10) logarithm of  {H$_2$ surface density assuming a Galactic CO-to-H$_2$ conversion factor}; 
  References for metallicity: $(a)$ \citet{Shi06}, $(b)$ \citet{C07}, $(c)$ \citet{Bego08}, $(d)$ \citet{Angel10}, $(e)$ \citet{Zhao10},
$(f)$ \citet{PMD03}, $(g)$ \citet{GdP03}. References for $\Sigma_{\rm H_2}$: $(h)$ \citet{Frayer98}, $(i)$ \citet{Leroy05}, $(j)$ \citet{Barone00}, $(k)$ \citet{albrecht04}, 
$(l)$ \citet{Leon98}, $(m)$ \citet{Verdes98}, $(n)$ \citet{Taylor98}, $(o)$ \citet{Tacconi&Young87}, $(p)$ \citet{Wei10}
\end{list}
\end{table*}

One goal of this paper is to investigate scaling relations between the CO emission and global properties of BCDs (e.g., luminosities, SFRs, metallicities). 
To this end, we have collected a large multiwavelength dataset from 
public archives and from the literature, as summarized in Table~\ref{T1.A1} and Table~\ref{T2.A1}.

We have supplemented our CO observations for \textit{subsample~I} with 
additional measurements for \textit{subsample~II} from different studies in 
the literature. 
With the aim of getting the more homogeneous dataset as possible, we have compiled 
the most recent CO\,(1$\rightarrow$0) fluxes, by giving a preference to those derived 
from observations at the IRAM 30~m telescope, if available. 
%

  
We have retrieved from the {\it GALEX} \citep{martin} 
{archive}\footnote{http://galex.stsci.edu/GalexView/} 
a fully reduced set of far ultra-violet (FUV: $\lambda 1530 \AA$) galaxy 
images from the All-Sky Imaging Survey (AIS). 
The angular resolution of the images is about 6$''$. 
Details on the data characteristics can be obtained from \citet{morrissey}. 
After background subtraction we have performed aperture photometry using 
circular apertures of 22$''$, i.e., {the} same as the CO (1$\rightarrow$0) beamsize. 
FUV luminosities have been corrected{for} galactic extinction, 
using the \citet{schlegel98} dust map and the \citet{cardelli} extinction 
curve, and {for} dust attenuation using the prescriptions of \citet{Buat05}, i.e., 
using the FIR to UV flux ratio.

{We use} $B-$band absolute magnitudes {from}
\citep{C01a}. 
Luminosities and sizes derived for both the BCD host galaxies and
their starburst region have been taken from the multi-band 2D surface
photometry presented by \citet{amorin09}. 
From their study we have also collected stellar masses, which have been 
derived using the luminosities and colours of the host galaxies and following  \citet{Bell&deJong01}. Typical uncertainties for stellar masses are below a 
factor of $\sim$\,2.

In addition, we {have collected} H$\alpha$ integrated luminosities $L_{\rm H\alpha}$ 
from \citet{C01b}, \citet{GdP03} and \citet{Angel10}, which have been
used to derive H$\alpha$-based SFRs.  
In the case of UM\,462 we have transformed H$\beta$ integral luminosities from \citet{Lagos07} to H$\alpha$ luminosities using the theoretical Balmer ratio 
assuming case B recombination with $T_e=$10$^4$\,K and $n_e=100$\,cm$^{-3}$.
For the SFRs, we have used the calibration given by \citet{Kennicutt09},

\begin{equation}\label{eq1} 
{\rm SFR(H\alpha)} = 7.9 \times 10^{-42} (L_{\rm H\alpha} + 0.0024 L_{\rm TIR}) [\rm{erg \ s}^{-1}]
\end{equation}
that accounts for internal dust attenuation using the total infra-red (IR)  
luminosity, $L_{\rm TIR}$ \citep{Dale2002}. 

The collected IR data consist in $K-$band magnitudes\footnote{In all cases the $K$ magnitudes are those obtained with the largest aperture ($\sim$4 times the $J-$band effective radius).} from the 2MASS extended source catalogue \citep[XSC;][]{Jarrett}, and 60$\mu m$ and 100$\mu m$ fluxes ($F_{60}$ y $F_{100}$) from the {\it IRAS} Faint Source Catalog \citep{Moshir}. 
The latter were obtained with a beamsize of 1.44$'$ and 2.94$'$ respectively, with a 
pointing error of about 30$''$ that was used as the search radius around the central target coordinates. Detections have medium or good quality for most of the galaxies, with  
uncertainties below $\sim$10\%, and only for few galaxies,  {namely} Mrk~324, Mrk~36, or III\,Zw~107 the uncertainties are between 20\% and 30\%. 
We calculated FIR and TIR fluxes and luminosities by following the prescriptions of 
\citet{Helou88} and \citet{Dale2002}. 

Furthermore, we use {the} H{\sc i} emission line fluxes, line widths 
and systemic velocities from \citet{TM81}, {and} \citet{G&G81} (only 
for Haro~1, Mrk~297 and III\,Zw~107). 
The velocity resolution of the data is 13 and 10 km s$^{-1}$, respectively.  
We have also used 1.4~GHz continuum luminosities derived from the 
NRAO VLA Sky Survey (NVSS) \citep{Condon98} total fluxes. 
The spatial resolution (FWHM) of the NVSS images is 45$''$ and
the sensitivity limit (rms noise level) is $\sim$0.45 mJy beam$^{-1}$. 


 %
\begin{figure*}[ht!]
\centering
\begin{tabular}{c}
\includegraphics[width=14.cm,angle=0]{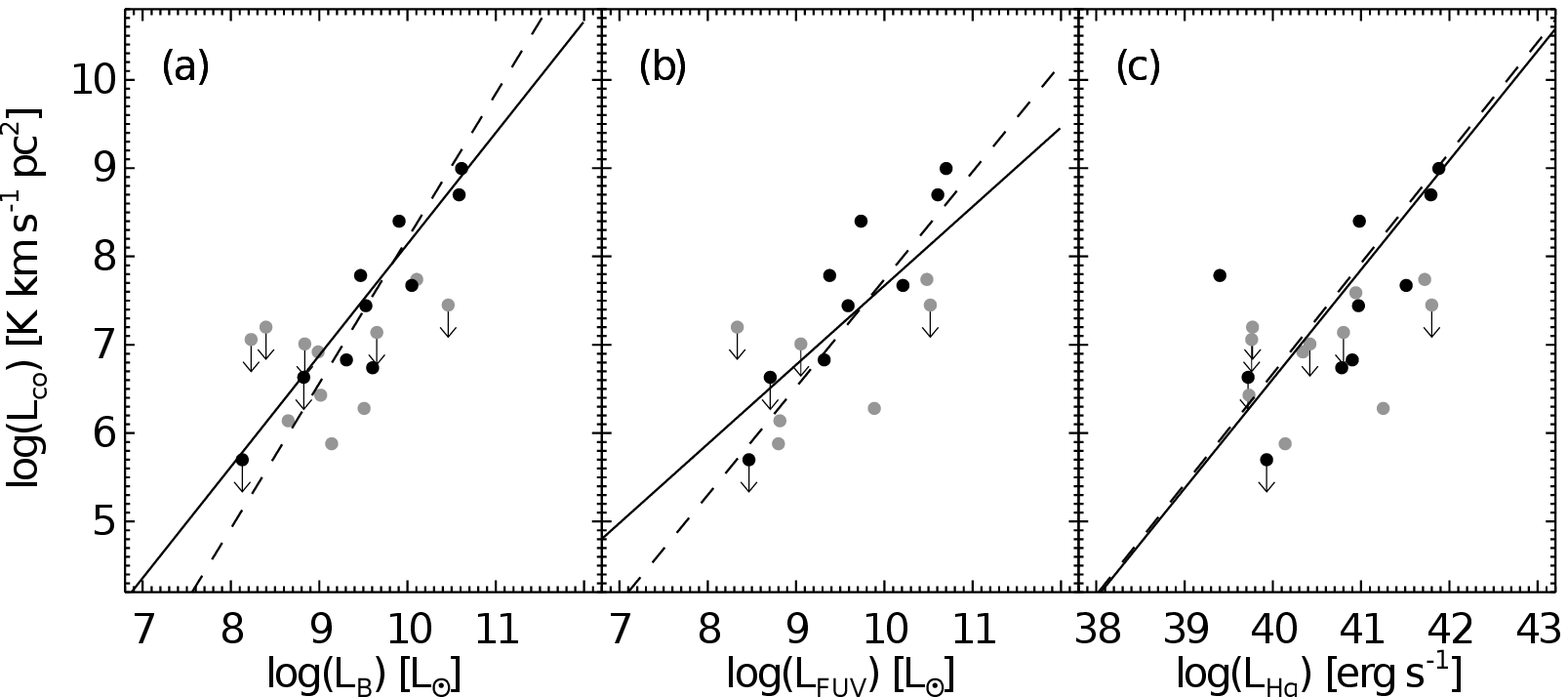}\\\vspace{0.5cm}
\includegraphics[width=14.cm,angle=0]{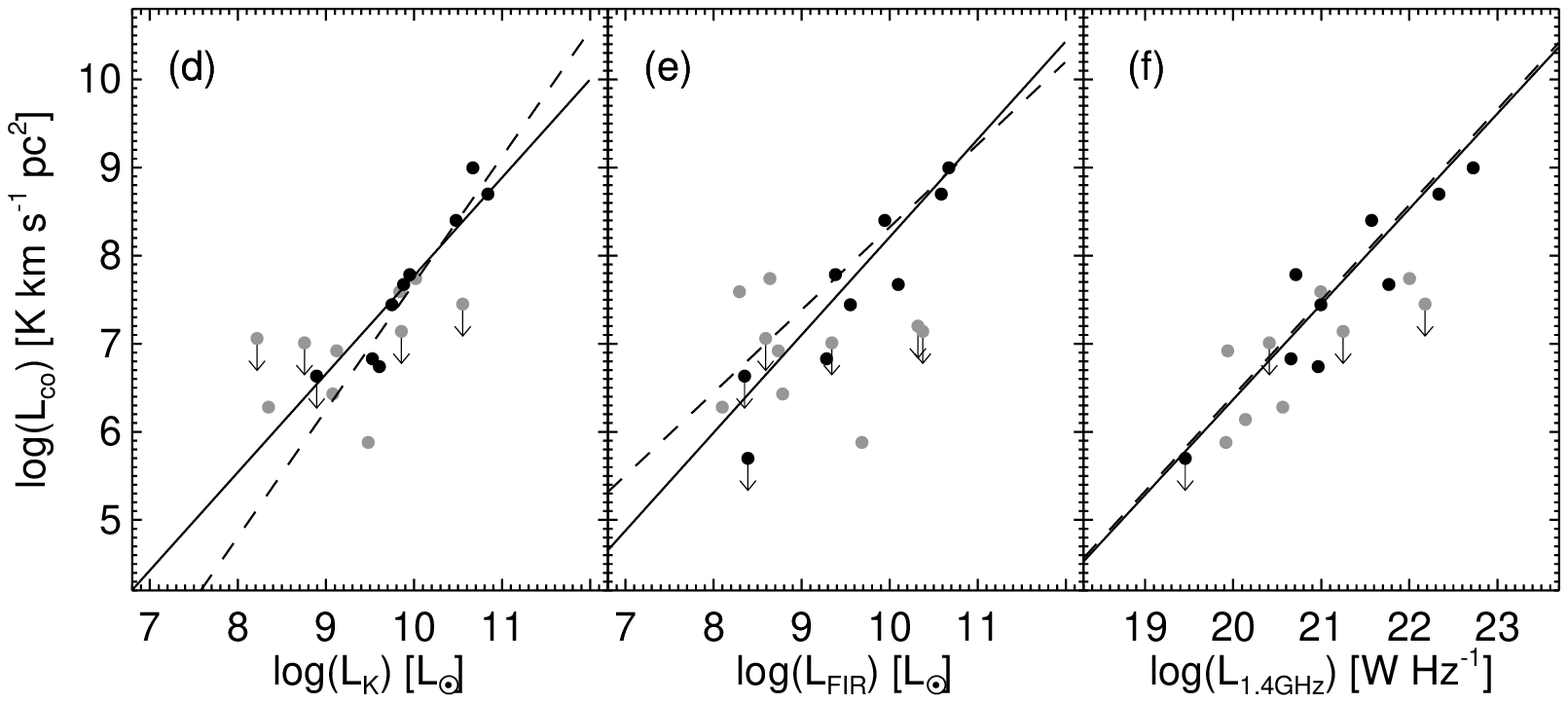}
\end{tabular}
\protect\caption{CO luminosity as a function of luminosity in
  different wavelength bands: $B$ $(a)$, $FUV$ $(b)$, H$\alpha$ $(c)$,
  $K$ $(d)$, $FIR$ $(e)$, and 1.4 GHz $(f)$. 
Black and grey points and lines show {\it subsample~I} and {\it
  subsample~II} BCDs, respectively. Arrows indicate CO upper limits.  
Dashed {and solid} lines indicate linear fits to detected only and all 
(detected and non-detected) galaxies, respectively.}\label{fig3}
\end{figure*}

Finally, we have collected from the literature gas-phase metallicities 
for the entire BCD sample. They were derived following the direct ($T_e$) method or 
strong-line methods calibrated through H{\sc ii} galaxies or Giant H{\sc ii} regions 
with good determination of the electron temperature, e.g., using the 
$N2$\,($\equiv$\,log([N{\sc ii}]/H$\alpha$)) index and the relations by \citet{PMC09}. 
Although this was not possible for the entire sample, for some extended BCDs 
(e.g., Mrk\,35, Mrk\,297, Mrk\,314, III\,Zw\,102 or Mrk\,401) instead
of the average (integrated) metallicity,  we have used the metallicity
measured in the star-forming regions covered by the CO observations. 
Taken into account the typical uncertainties associated with different
metallicity calibrations and methods 
we have considered an average uncertainty value of $\sim$0.2 dex. 

\section{CO emission in Blue Compact Dwarfs}  
\label{s4}

\subsection{Detections and line measurements}
\label{s4.1}

In Table~\ref{T2} we summarize the results from the
CO observations, including values for the 
CO line velocities, line widths, and integrated fluxes. 
We detect ($\geq$\,3$\sigma$) CO emission in seven out of ten 
BCDs: Mrk\,33, Mrk\,35, Mrk\,297, Mrk\,401, Haro\,1, III\,Zw~102, and
III\,Zw~107. Two of them, Mrk\,401 and III\,Zw~107, did not presented
previous CO detections in the literature\footnote{Based on the NED:  http://ned.ipac.caltech.edu/}. {For III\,Zw~107  
we only have 3$\sigma$ detection at 230~GHz and a marginal detection 
($<$\,3$\sigma$) at 115~GHz after smoothing the spectra to 20 km\,s$^{-1}$. 
For the three remaining galaxies, Mrk\,36, and Mrk\,324 are not
detected, while
 Mrk~314 shows a tentative $\sim$\,2$\sigma$ detection only at
 230~GHz, with 
fluxes in good agreement with previous IRAM observations \citep{Garland2005}}. 
The spectra of the galaxies detected in CO are presented in 
Fig.~\ref{fig2a}, while spectra of marginally or non-detected galaxies 
are shown in Fig.~\ref{fig2b}.

For each detection a Gaussian function was fitted to the CO emission line 
to determine the central velocity $V_{\rm co}$, the velocity width at
half {maximum} $\Delta V_{\rm co}$, and the integrated flux 
$I_{\rm CO} = \int{T_{\rm mb} \ \delta\nicev}$. 
For those BCDs showing more than one velocity component, 
e.g., Mrk\,33 or Mrk\,35, we fitted up to two Gaussians. 
For galaxies showing clear asymmetries in the CO line profile, e.g., 
Mrk\,297 or I\,Zw\,102, we have also estimated $I_{\rm co}$ by the 
numerical integration of the smoothed spectra within the limits 
set by the line. 
In all cases, line flux uncertainties were estimated following 
\citet{elfhag} and \citet{albrecht04}. 
Upper limits for non-detections have been estimated as $I_{\rm co}$\,
$\la$\,3$\sigma$\,$\sqrt{\Delta V_{\rm CO} \ \delta\nicev}$, where 
$\sigma$ is the rms noise level obtained in the baseline range, 
$\Delta$\,$V_{\rm co}$ is the expected total velocity width of the 
CO line and $\delta\nicev$ is the channel velocity width.  
For non-detections we have assumed $\Delta V_{\rm co}$ as the total 
velocity width of the H{\sc i} profile. 
To check the validity of this approximation we have confirmed that, 
for clearly detected galaxies, the ratio between CO and H{\sc i} line 
widths is always below one, regardless of the number of Gaussian 
components.

\subsection{CO luminosity}
\label{s4.2}

\begin{table*}[ht!]
\caption{CO measurements}
\label{T2}
\centering
\begin{tabular}{l c c c c c c c c c c c c c}
\noalign{\smallskip}
\hline\hline
\noalign{\smallskip}
Galaxy & $V_{\rm co}$ & $\Delta V_{\rm co}$ & $T_{\rm A, 1-0}$ & $T_{\rm A*, 2-1}$ & 
$\delta T_{rms, 1-0}$ & $\delta T_{rms, 2-1}$ & $I_{\rm co, 1-0}$ & $I_{\rm co, 2-1}$ 
& log\,$L_{\rm co}$& $\log \Sigma_{\rm H_2}$  \\ 
& km s$^{-1}$ & km s$^{-1}$ & mK & mK & mK & mK & K km s$^{-1}$ & K km s$^{-1}$ 
&  K km s$^{-1}$ pc$^2$ & M$_{\odot}$\,pc$^{-2}$ \\
(1)  & (2) & (3) &  (4) & (5) & (6) & (7) & (8) & (9) & (10) & (11) \\ 
\noalign{\smallskip}
\hline
\noalign{\smallskip}
Haro\,1 & 3793.0$\pm$6 & 151 & 73.6 & 76.0 & 8.9 & 12.0 & 15.1$\pm$0.5 & 19.4$\pm$0.8 &  8.7 & 1.81 \\[3pt]

Mrk\,33 & 1482.5$\pm$8 & 85  & 43.5 & 46.3 & 7.0 & 4.9 & 4.6$\pm$0.3 & 6.7$\pm$0.5 &  7.4 & 1.14 \\[3pt]

Mrk\,35 & 939$\pm$6    & 90  & 23.6 & 37.0 & 5.0 & 9.2 & 2.3$\pm$0.2 & 2.0$\pm$0.3 & 6.8 & 0.85 \\[3pt]

Mrk\,36 & 646$\pm$7    & ... & ...  & ...& 4.7 & 5.0 &  $<$0.4 & $<$0.6 & $<$5.7 &  $<$0.10 \\[3pt] 

Mrk\,297& 4731$\pm$5   & 127 & 97.7 & 115 & 7.2 & 9.2 & 19.2$\pm$0.6 & 27.3$\pm$0.2 & 9.0 & 1.81\\[3pt] 

Mrk\,314& 2091$\pm$6$^a$   & 161$^a$& ...  & 5.4 & {5.6} & 2.7 & {$<$0.6} & 1.7 & 6.7$^a$ & 0.13$^a$\\[3pt]

Mrk\,324& 1600$\pm$6   &...  &  ... & ... & 5.6 & 5.3 & $<$0.7 &$<$0.9 & $<$6.6  & $<$0.39\\[3pt]

Mrk\,401& 1699$\pm$6   & 43  & 148.6 & 169.1 & 6.6 & 9.1 & 8.6$\pm$0.3 & 15.2$\pm$0.5 & 7.8 & 1.46 \\[3pt]   
     
III\,Zw\,102& 1615$\pm$4&130  & 203.7 & 201.4 & 15.3 & 20.7 & 40.0$\pm$0.4 & 38.3$\pm$0.5 & 8.4 & 2.23 \\[3pt]

III\,Zw\,107 & 5785$\pm$11 & 120 & 3.7 & {7.8} & {2.0} & {2.6} & {0.6$\pm$0.3} & {0.7$\pm$0.2} &  {7.7$^a$} & {0.24$^a$}\\[3pt]
\noalign{\smallskip}			
\hline									
\hline	    							
\end{tabular}						
\begin{list}{}{}
\item Notes.$-$ Columns: (2) and (3) CO 1$\rightarrow$0 line central
  LSR velocity and width at half maximum; (4)-(5) and (6)-(7) {peak} intensity
  and rms of CO 1$\rightarrow$0 and CO 2$\rightarrow$1 lines, respectively; 
  (8) and (9) {integrated} line emission and 1$\sigma$ error bars in $T_{\rm mb}$ units 
  ($I_{\rm co}$ $= \frac{F_{\rm eff}}{B_{\rm eff}} \int T_{\rm A}^{\star}$dv).  
  3$\sigma$ upper limits are given for non-detections; (10) and (11) CO 
  (1$\rightarrow$1) luminosity and H$_2$ surface brightness  {assuming
    a Galactic CO-to-H$_2$ conversion factor}. {$^a$\,It corresponds to
    the 2$\rightarrow$1 line, which is detected at
    $\sim$\,3$\sigma$}. 
Its integrated flux was converted to CO 1$\rightarrow$0 flux to derive the values in (10) and (11). 
\end{list}	
\end{table*}		
	
We derive the CO luminosity of the BCD sample as 
$L_{\rm CO} =$ 23.5 $\Omega_{\rm obs}$ $D^{2}_{L}$ $I_{\rm CO}$
$(1+z)^{-3}$ [K km s$^{-1}$ pc$^{2}$] \citep{solomon97}, where $I_{\rm
  co}$ refers to the CO\,(1$\rightarrow$0) integrated fluxes,
$\Omega_{\rm obs}$ is the observed solid 
angle at the given frequency and $D_{L}$ is the luminosity distance of the galaxy. 
The $L_{\rm CO}$ values for the observed galaxies  {have been listed} in Table~\ref{T2}.
 
\begin{figure}[t!]
\centering
\includegraphics[width=8.5cm,angle=0]{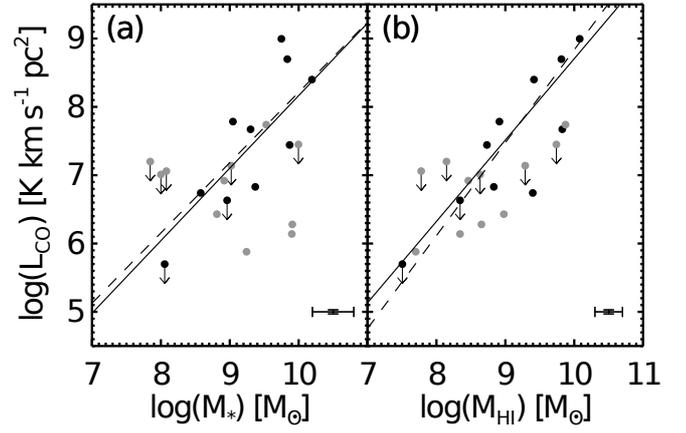}
\protect\caption[Masses]{CO luminosity as a function of stellar mass 
$(a)$ and H{\sc i} gas mass $(b)$. 
Symbols and colours are as in Fig.~\ref{fig3}. 
}    
\label{fig4}
\end{figure}

\begin{figure}[t!]
\centering
\includegraphics[width=8.5cm,angle=0]{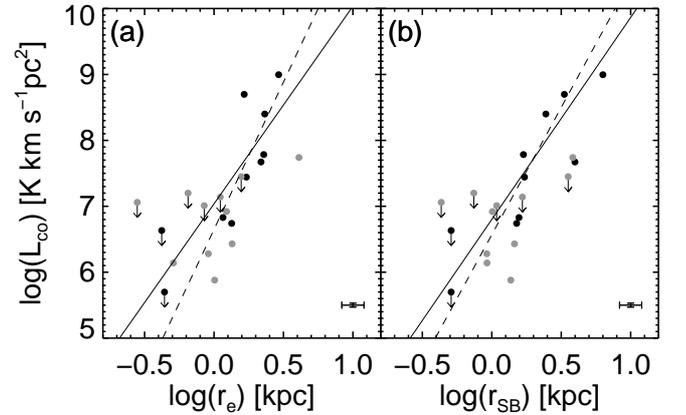}
\protect\caption[Sizes]{Relation between CO luminosity and effective
  radius of the host galaxy $(a)$, and the size of the starburst
  emission in the optical $(b)$. 
Symbols and colours are as in Fig.~\ref{fig3}.}    
\label{fig5}
\end{figure}

\begin{figure}[ht!]
\centering
\includegraphics[width=8.cm,angle=0]{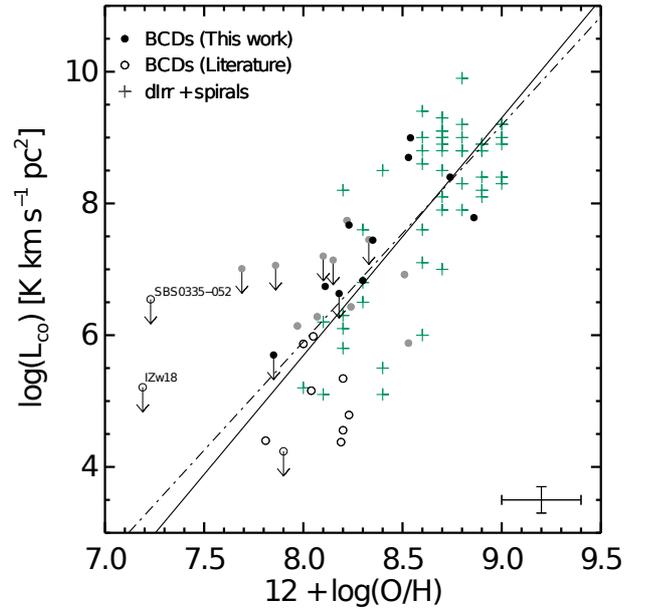}
\protect\caption[]{CO luminosity as a function of metallicity.  
Black dots show {\it subsample~I} while open circles show {\it subsample~II},
and additional very low-metallicity BCDs compiled from the literature \citep{Fumagalli10}. 
Green crosses show the compilation of nearby star forming disc
galaxies in \citet{Krumholz09}. Lines indicate linear best-fits to all
galaxies (solid) and BCDs only (dashed). {Error bars indicate typical
  uncertainties for our sample.}}    
\label{fig6}
\end{figure}

From Fig.~\ref{fig3} to Fig.~\ref{fig7} we investigate scaling relations  
between $L_{\rm CO}$ and other key properties that are directly
connected to star formation activity, as well as stellar and gas content in galaxies. 
In particular, we use luminosities at different wavelengths such as 
$L_{\rm FUV}$, $L_{\rm B}$, $L_{\rm K}$, $L_{\rm FIR}$, $L_{\rm 1.4 GHz}$ and 
$L_{\rm H\alpha}$, stellar ($M_{\rm star}$) and total H{\sc i} gas ($M_{\rm HI}$) 
masses, effective radius of the host galaxies ($r_e$) and projected sizes of 
the starburst region ($r_{\rm SB}$), as well as the gas-phase 
metallicity. 
Linear least-square fits to data in Figs.~\ref{fig3}-\ref{fig7} 
were performed using the routine {\sc fitexy} \citep{Press92}. 
The results 
are presented in Table~\ref{correlations}.

In Fig.~\ref{fig3} we show the tight correlation found 
between the CO luminosity and the luminosities in several 
{wavelength} bands, most of them classical tracers of the 
SFR in galaxies. 
Overall, we find that less luminous BCDs are also fainter in CO.  
The fitted relations have slopes between $\sim$0.9-1.5. 
The derived Spearman rank indices $\rho$ 
(see Table~\ref{correlations}) indicate tight correlations between the  
luminosities considered. The only exception is $L_{\rm FIR}$, which
shows a large scatter and a lower $\rho$. 
The statistical errors provided by the least-square fitting
are typically {a} few \%. 

In Fig.~\ref{fig4} we also find {tentative} correlations between the CO 
luminosity and the total H{\sc i} gas mass and stellar mass of the BCD hosts. 
The latter is in relatively good agreement with the $L_{\rm
  CO}$-$L_{\rm K}$ relation, as it is expected since the $L_{\rm K}$ is a proxy of stellar mass. 
Between these two, the relation between $L_{\rm CO}$ and $M_{\rm
  star}$ show a larger scatter, possibly due to larger uncertainties
in the derivation of $M_{\rm star}$, i.e., in the colours of the host 
galaxy and given by several assumptions done by models in the derivation 
of the mass-to-light ratios \citep[e.g., initial mass function or the star
formation history,][]{amorin09}.
			
In Fig.~\ref{fig5} we find that the CO luminosity also scales with
size. The correlations between both the 
effective radius of the host galaxy ($r_{\rm e}$) and the optical size 
of their starburst ($r_{\rm SB}$) indicate that larger BCDs, which
have extended and luminous starbursts \citep{amorin09}, are best detected 
in CO than smaller BCDs with more compact star-forming regions.

Finally, in Fig.~\ref{fig6} we show the CO luminosity-metallicity 
relation. In addition to our galaxy sample we also include in
Fig.~\ref{fig6} CO luminosities for several nearby late-type  
irregular and spiral galaxies from the literature, including upper 
limits for the most extremely metal-poor BCDs known, I~Zw~18 
\citep{Vilchez98,Leroy07} and SBS~0335-052 \citep{Dale01,P06}. 
Despite the large scatter, we find a strong positive correlation 
showing the CO luminosity rapidly increasing with metallicity. 
The same slope is found if we use $I_{\rm co}$ instead of $L_{\rm
  co}$, in 
agreement with previous studies \citep[e.g.,][]{Taylor98,Schruba12}. 

\subsection{Molecular mass and surface densities}
\label{s4.3}

The molecular hydrogen mass and surface density {\it traced} by CO 
can be estimated from CO luminosities. 
Unfortunately, this implies to assume a CO$-$to$-$H$_2$ conversion 
factor, $X_{\rm CO}$, which strongly depends on the physical conditions 
of the gas \citep[][for an extended review]{Bolatto13}. 
A frequently used $X_{\rm CO}$ value is obtained from Milky Way virialised
molecular clouds. For this reason their use in other galaxies 
must be taken with caution. 
This is especially true in metal-poor galaxies, 
since, as we will discuss later, several observational studies have found 
$X_{\rm CO}$ as a strong function of metallicity 
\citep[e.g.,][]{Rubio93,Wilson95,Arimoto96,Israel97,Leroy11,Genzel12,Schruba12,Bolatto13}.  
Keeping this in mind, we have adopted as a first order approach a constant 
CO-to-H$_2$ conversion factor 
$X_{\rm CO, MW}$\,$=$\,2$\times$10$^{20}$ cm$^{-2}$ (K km s$^{-1}$) or
$\alpha_{\rm CO, MW}$\,$=$\,3.2\,M$_{\odot}$ (K km s$^{-1}$ pc$^2$)$^{-1}$ 
 \citep{Strong&Mattox96,Dame01}, which is roughly the mean of values 
estimated in the Milky Way and nearby galaxies.  
Including a factor of 1.36 due to the contribution of {helium}, 
the total molecular mass traced by CO (in solar units) and the CO
surface density (in units of M$_{\odot}$\,pc$^{-2}$)  
were then derived as $M_{\rm H_2}$\,$=$\,1.36\,$L_{\rm
  CO}$\,$\alpha_{\rm CO, MW}$ and 
$\Sigma_{\rm H_2}$\,$=$\,1.36~$\cos$\,$i$~$\alpha_{\rm CO, MW}$~$I_{\rm CO}$, 
where $i$ is the inclination of the galaxies (see Table~\ref{T1.A1} and Table~\ref{T2.A1}). 

%

Using the above prescriptions, our sample of BCDs show a wide range of 
CO masses, 
M$_{\rm H_2}$\,$\sim$\,2$\times$10$^6$--\,4$\times$10$^9$\,M$_{\odot}$,  
and CO surface densities, 
$\Sigma_{\rm H_2}$\,$\sim$\,1--170\,M$_{\odot}$\,pc$^{-2}$. 
Most of these values are typical for late-type irregular and spiral 
galaxies \citep[cf. e.g.,][]{K98,Leroy05}, except for the three
luminous {BCDs} 
of the sample, which are clearly more massive and dense in CO.
\begin{table*}[ht!]
\caption{{Correlations}}
\label{correlations}
\centering
\begin{tabular}{c c c c c c}
\noalign{\smallskip}
\hline\hline
\noalign{\smallskip}
$x$ & $y$ & Sample & $a$ & $b$ & $\rho$  \\ 
(1)    &   (2) &    (3)     &   (4)  & (5) & (6)  \\ 
\noalign{\smallskip}
\hline
\noalign{\smallskip}
$\log(L_{\rm B})$&$\log(L_{\rm CO})$& det & 1.64$\pm$0.01 & -8.20$\pm$0.08 &  0.77      \\\vspace{1mm}
&& all   &   1.26$\pm$0.01 & -4.46$\pm$0.04 &  0.71 \\\hline \noalign{\smallskip}

$\log(L_{\rm FUV})$&$\log(L_{\rm CO})$& det & 1.22$\pm$0.02 & -4.46$\pm$0.14 &  0.76     \\\vspace{1mm}
&& all & 0.89$\pm$0.01 & -1.28$\pm$0.09 &  0.75     \\\hline \noalign{\smallskip}

$\log(L_{\rm H\alpha})$&$\log(L_{\rm CO})$& det &  1.25$\pm$0.01 & -43.4$\pm$0.36 & 0.70  \\\vspace{1mm}
&& all   &  1.24$\pm$0.01& -42.9$\pm$0.20 & 0.79  \\\hline \noalign{\smallskip}

$\log(L_{\rm K})$&$\log(L_{\rm CO})$& det   &  1.44$\pm$0.01 & -6.70$\pm$0.07  &   0.94    \\\vspace{1mm}
&& all   &  1.11$\pm$0.01 &  -3.38$\pm$0.05  &   0.83    \\\hline \noalign{\smallskip}

$\log(L_{\rm FIR})$&$\log(L_{\rm CO})$& det   &  0.94$\pm$0.01  &  -1.07$\pm$0.07 &  0.57   \\\vspace{1mm}
&& all   &  1.11$\pm$0.01  &   -2.90$\pm$0.05  &  0.57     \\\hline \noalign{\smallskip}

$\log(L_{\rm 1.4GHz})$&$\log(L_{\rm CO})$& det   &  1.08$\pm$0.01  &  -15.3$\pm$0.1  & 0.85  \\\vspace{1mm}
&& all   &  1.08$\pm$0.01  &  -15.3$\pm$0.1  & 0.85  \\ \hline \noalign{\smallskip}

$\log($M$_{*})$  & $\log(L_{\rm CO})$& det & 1.03$\pm$0.01 & -2.08$\pm$0.06 & 0.59 \\\vspace{1mm}
&& all & 1.06$\pm$0.01 & -2.45$\pm$0.04 & 0.40  \\ \hline \noalign{\smallskip}

$\log($M$_{\rm HI})$  & $\log(L_{\rm CO})$& det & 1.36$\pm$0.01 & -4.76$\pm$0.08 & 0.77  \\\vspace{1mm}
&& all & 1.19$\pm$0.01 & -3.20$\pm$0.04 & 0.72  \\ \hline \noalign{\smallskip}

$\log(r_{\rm e})$  & $\log(L_{\rm CO})$& det   & 4.47$\pm$0.01 &   6.65$\pm$0.03 & 0.86  \\\vspace{1mm}
&& all   &  2.99$\pm$0.01 &  7.03$\pm$0.01  & {0.74}  \\ \hline \noalign{\smallskip}

$\log(r_{\rm SB})$  & $\log(L_{\rm CO})$& det   & 3.86$\pm$0.01 &  6.57$\pm$0.02 & 0.86  \\\vspace{1mm}
&& all   & 3.06$\pm$0.01  &  6.80$\pm$0.01 & 0.77  \\ \hline \noalign{\smallskip}

$12+\log(O/H)$&$\log(L_{\rm CO})$& BCD   &   3.80$\pm$0.03 & -24.9$\pm$0.20 &  0.76   \\
&& all   &   3.23$\pm$0.02  &  -20.1$\pm$0.17  &  0.77    \\ \hline
\noalign{\smallskip}

$12+\log(O/H)$& $\log(\Sigma_{\rm H_2}/\Sigma_{\rm HI})$ & det & 2.38$\pm$0.32 &  -20.13$\pm$2.74 & 0.85 \\
&& all & 1.98$\pm$0.23 &  -16.75$\pm$1.88 &0.87  \\ \hline
\noalign{\smallskip}

$12+\log(O/H)$& $\log(\Sigma_{\rm H_2}/\Sigma_{\rm H_2+HI})$ & det & 1.29$\pm$0.32 &  -11.28$\pm$2.74 &  0.87\\
&& all & 1.17$\pm$0.23 &  -10.30$\pm$1.88 & 0.85 \\ \hline
\noalign{\smallskip}

$12+\log(O/H)$& $\log(\Sigma_{\rm SFR}/\Sigma_{\rm H_2})$& det & -1.91$\pm$0.32 &  17.00$\pm$2.74 & -0.92 \\
&& all & -1.43$\pm$0.23 &  12.92$\pm$1.88 & -0.82 \\ \hline
\noalign{\smallskip}

$12+\log(O/H)$& $\log(\Sigma_{\rm SFR}/\Sigma_{\rm H_2+HI})$& det & -0.62$\pm$0.32 &  5.72$\pm$2.73 & -0.33 \\
&& all & -0.26$\pm$0.23 &  2.62$\pm$1.88 & -0.17 \\

\noalign{\smallskip}			
\hline									
\end{tabular}								
\begin{list}{}{}
\item Notes.$-$ Columns (4) and (5) show the best-fit coefficients, slope
  $a$ and intercept $b$ with their 1$\sigma$ dispersion, {in the relations ($y = ax + b$)} 
  between quantities in Cols. (1) and (2). 
  Column (3) indicates the sample of galaxies considered to make the 
  linear fits: galaxies including detections and upper limits (all), and only 
  detected galaxies (det). {Column (6)} lists the Spearman rank index 
  $\rho$. 
\end{list}	
\end{table*}									
\section{Star formation laws}
\label{s5}
 {In this section we study the relation between star formation rates and gas 
(molecular and neutral) content averaged over the entire starburst region. 
In addition to the natural interest of further comparing the results provided by 
studies on sub-kpc scales, global studies of BCDs can provide an interesting 
benchmark for future comparison studies with analogues at higher redshifts, for which 
we can only measure properties at kpc-size scales even using interferometry.}

We have explored the position of the BCD sample in the $\Sigma_{\rm SFR}$-$\Sigma_{\rm gas}$ 
plane and  {compared} with the SK law  {followed by normal galaxies and 
luminous starbursts.}
For that purpose we have derived surface densities in H{\sc i}  
($\Sigma_{\rm HI}$), SFR ($\Sigma_{\rm SFR}$) and total H{\sc i}$+$H$_2$ gas 
($\Sigma_{\rm gas}$\,$=$\,$\Sigma_{\rm HI+H_2}$). 
We have followed \citet{K98} and adopted a circular
aperture for each galaxy 
that enclose the region where the star-forming regions are distributed. 
Thus, for the size of each aperture we have used alternatively 
$r_{\rm 25}$, $r_{\rm SB}$ or the CO beam size after correction 
for inclination.
For the star formation rate we have considered the {\it ongoing} SFR,
as given by the dust-corrected H$\alpha$ luminosity. 
{The uncertainties for the SFR and gas surface densities are 
mainly dominated by 
possible galaxy to galaxy variations in the apertures considered.  
For the following analysis, we acknowledge a conservative, average 
uncertainty of $\sim$\,0.3 dex for surface densities, which do not
account by uncertainties in the CO-to-H$_2$ conversion factor and by
assumptions involved in the derivation of SFR (e.g., the tracer and
IMF in use or the adopted SF history)}. 

\begin{figure*}[ht!]
\centering
\includegraphics[width=15.cm,angle=0]{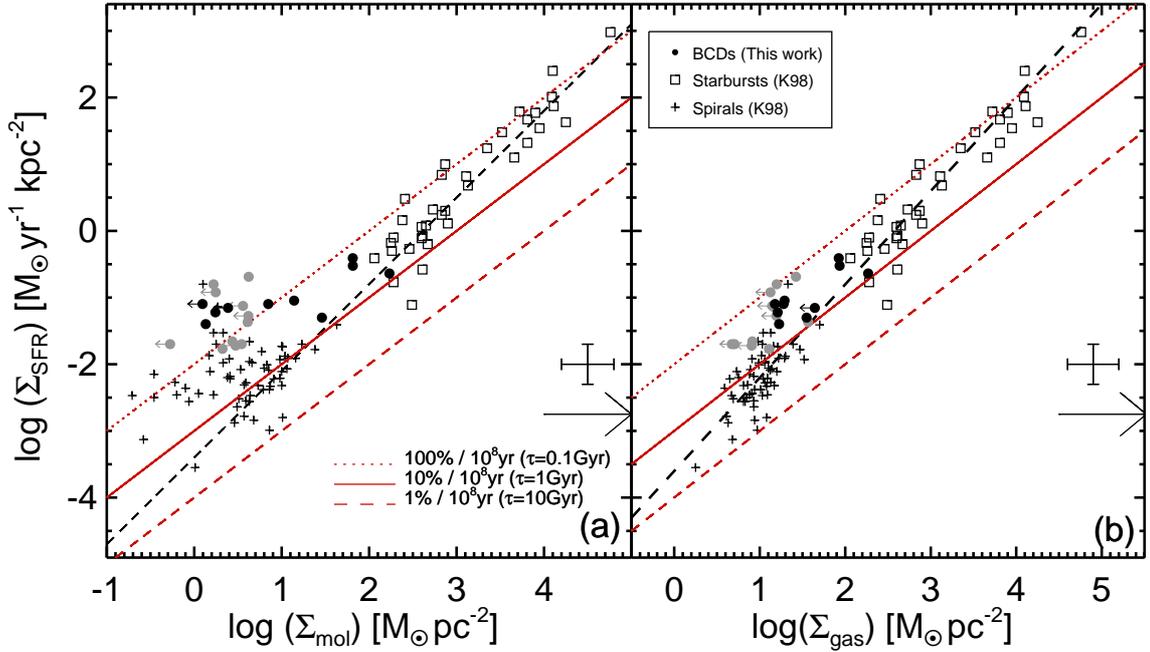}
\protect\caption[SK laws]{ {Star formation rate surface density as a
  function of H$_2$ (\textit{Left}) and H{\sc i}$+$H$_2$ (\textit{Right}) gas 
  surface densities. Black dots show {\it subsample~I} and grey dots show 
  {\it subsample~II}. 
The black dashed lines show the fits to the H$_2$ and H$_2+$H{\sc i} SK law by
\citet{Leroy05} and \citet{Shi11}, respectively. 
Red lines {indicate} constant star formation efficiencies (depletion timescales). 
Large arrows indicate the shift in H$_2$ surface density by assuming a 
CO-to-H$_2$ conversion factor {10 times} larger than that of the 
Milky Way for the molecular mass. {Error bars indicate typical
  uncertainties for our sample.}}}    
\label{fig7}
\end{figure*}

In Figure~\ref{fig7} we present the BCD SFR surface density as a 
function of their H$_2$ and H$_2+$H{\sc i} surface densities. 
In {this figure} we have also included the data (spiral discs 
and starburst galaxies) used by \citet{K98} for the calibration of the SK law. 
Instead of including the fit by \citet{K98}, we have included 
in Fig.~\ref{fig7}$a$ a slightly different fit 
($\Sigma_{\rm SFR}$\,$=$\,10$^{-3.4}$\,$\Sigma_{\rm H2}^{1.3}$) found by
\citet{Leroy05} for a large sample of dwarf irregular galaxies and normal spirals.  
Similarly, in Fig.~\ref{fig7}$b$ we have included the fit presented by 
\citet{Shi11} ($\Sigma_{\rm SFR}$\,$=$\,10$^{-3.9}$\,$\Sigma_{\rm H_2+HI}^{1.4}$) 
for a largest sample of late-type, early-type, and
starburst galaxies at low and high redshift, which includes the sample
of galaxies studied by \citet{K98}. The relation of \citet{Shi11} is
also in excellent agreement with the more recent update of the SK relation by
\citet{Kennicutt12}. 

Overall, Fig.~\ref{fig7} shows the BCD sample located in a region 
corresponding to depletion timescales for both molecular and molecular plus 
atomic gas {of less than} 1 Gyr, showing surface densities in
{smaller than} starbusts 
and {larger than} spiral and irregulars discs. 
Remarkably, for most BCDs in our sample the depletion timescales appear 
to be up to $\sim$\,2 dex lower than expected from the SK-laws.
In {particular, Fig.~\ref{fig7}$a$ shows that most BCDs have} lower $\Sigma_{\rm H_2}$ 
than the one predicted by the \citet{Leroy05} relation 
for a given $\Sigma_{\rm SFR}$. 
Consequently, their H$_2$ depletion timescales are extremely short ($<$\,0.1 Gyr).  
Considering the total gas (H$_2+$H{\sc i}) surface density, 
most BCDs still show a systematic departure of up to $\sim$\,1 dex
from the SK relation to low $\tau_{\rm H_2+HI}$ that is slightly larger
than observational uncertainties, as shown in Fig.~\ref{fig7}$b$. 

Large offsets from the SK law for metal-poor star forming galaxies 
have been reported in previous studies. Rather than galaxies with
 {enhanced SF efficiency} or following a distinct SF law, these offsets 
have been attributed to changes in the CO-to-H$_2$ conversion 
factor at low metallicities \citep[e.g.,][]{Kennicutt12}.  
If instead of using a Galactic conversion factor $\alpha_{\rm CO, MW}$ 
we adopt a metallicity-dependent $\alpha_{\rm CO}$ value \citep[e.g.,][]{Arimoto96},  
those BCDs with strongly sub-solar metallicities would be displaced 
to higher gas surface densities and larger depletion timescales
({large} arrows in Fig.~\ref{fig7}). 
Additional support to this interpretation comes from the most luminous galaxies 
of the sample, Mrk\,297, III\,Zw\,102, and Haro\,1, which are the only BCDs 
that seem to follow the SK relations. 
These galaxies show significantly higher gas and SFR densities, placing them 
at the lower end of the trend followed by the IR-selected starbursts of 
\citet{K98}. {Up to} some point this is not surprising, since these objects are 
luminous in IR wavelengths, showing clear dust patches/lanes and 
complex/distorted morphologies in their inner regions 
\citep[][see also our Fig.~\ref{fig1}]{C01a,C01b} compatible with past or recent
mergers. Perhaps most importantly, these galaxies are the most metal-rich 
starbursts in the sample. 

In the following sections we discuss in detail how  {metallicity
  affects the molecular gas depletion timescales in  {the
    extreme ISM conditions (i.e., high ionization and high {\it
      specific} SFR)} of BCDs. This will lead us to find a method for} 
deriving a metallicity-dependent form of $\alpha_{\rm CO}$ that  {in turn allows 
us to derive corrected H$_2$ masses and revise our scaling relations} accordingly.   

\section{Gas depletion timescales and metallicity}
\label{s6}

\begin{figure*}[ht!]
\centering
\includegraphics[width=15.cm,angle=0]{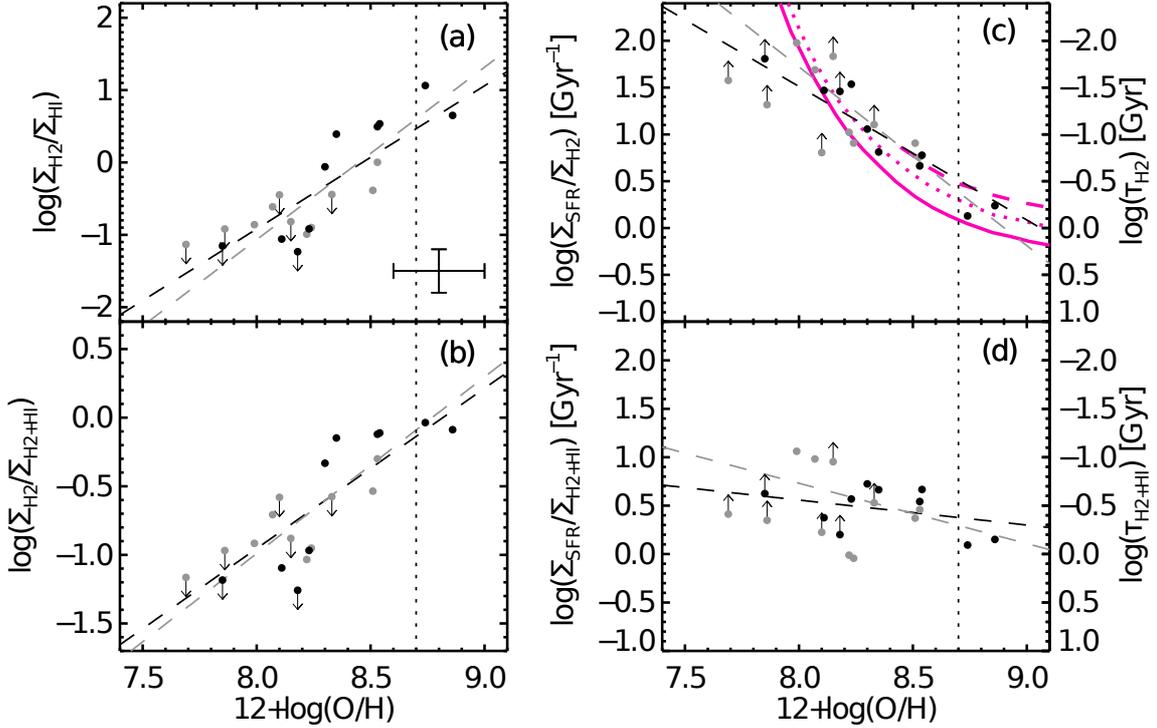}
\protect\caption[]{ {The molecular to atomic ratio $(a)$, the molecular fraction 
$(b)$, and the H$_2$ and H$_2+$H{\sc i} depletion timescale ($c$ and $d$, respectively) 
as a function of metallicity. Symbols are as in Fig.~\ref{fig7}. 
{Black and grey} dashed lines show linear best-fits {to all data
points and only secure detections, respectively}. 
Solid, dotted and dashed magenta lines in panel $c$ show model predictions by 
\citet{Krumholz11} for different values of $\Sigma_{\rm H_2+HI}$ (80, 20 and 5  
M$_{\odot}$\,pc$^{-2}$, respectively).}}    
\label{fig8}
\end{figure*}

In Figure~\ref{fig8} we investigate as a function of metallicity the 
following quantities:  
$(a)$ the molecular to atomic surface density ratio 
(i.e., $R_{\rm H_2}$\,$=$\,$\Sigma_{\rm H_2}/\Sigma_{\rm HI}$), 
$(b)$ the molecular to total (H$_2+$H{\sc i}) gas surface density ratio 
(i.e., the molecular fraction $f_{\rm H_2}$\,$=$\,$\Sigma_{\rm H_2}/\Sigma_{\rm H_2+HI}$) 
and, $(c)$ and $(d)$ the molecular and total gas (H$_2+$H{\sc i}) SFE 
and their inverse quantity, the gas depletion timescale 
$\tau_{\rm H_2}$ and $\tau_{\rm H_2+HI}$, respectively. 
 {The coefficients of the best-fit relations shown in Figure~\ref{fig8} 
are presented in Table~\ref{correlations}.} 

 {In Figures~\ref{fig8}\,$a-b$ we show that lower $R_{\rm H_2}$ and 
$f_{\rm H_2}$ are found for BCDs of increasingly lower metallicities}.   
On the other hand,  {in Fig.~\ref{fig8}$c$ we find a strong anticorrelation}, 
showing a fast decrease of the SFE from low- to high-metallicity BCDs. 
The latter implies that the H$_2$ depletion timescale in BCDs is an increasing 
function of metallicity, showing a large variation in $\tau_{\rm H_2}$ 
 {(up to a factor of $\sim$\,50)}, from $\sim$0.02 Gyr for BCDs with 
$Z$\,$\sim$\,0.1\,$Z_{\odot}$ to $\sim$1 Gyr for BCDs with nearly solar abundance. 
 {The relation found between $\Sigma_{\rm SFR}/\Sigma_{\rm H_2}$ and 
metallicity in Fig.~\ref{fig8}\,$c$ {appears} in good qualitative agreement with 
model predictions by \citet[][magenta lines, see also Krumholz et al. 2009]{Krumholz11}.}
Finally, the dependence of the SFE on metallicity is significantly weaker when we 
consider the total gas surface density, as shown in Fig.~\ref{fig8}$d$. 

 {Short depletion timescales for H$_2$} in low-mass spirals and nearby 
dwarf galaxies have been previously reported in the literature \citep[e.g.,][]{Leroy06,Leroy07}. 
Although, on average, molecular gas in nearby disk galaxies is consumed 
uniformly, i.e., the efficiency with which molecular gas {forms} stars 
does not depend on the surface density of H$_2$ averaged over large scales  
\citep{Bigiel08,Bigiel11,Leroy08,Schruba11}, variations in the SFE among galaxies 
are usually found. Thus, least massive galaxies tend to show shorter depletion 
timescales for molecular gas than more massive systems 
{\citep[e.g.,][]{Gratier2010a,Schruba11,Saintonge11b,Leroy13}.}
 
In star-forming dwarf galaxies, where the range of metallicities is large, 
i.e., from nearly solar to few per cent solar, H$_2$ depletion timescales are 
generally lower than $\sim$\,2 Gyr, the galaxy-averaged value found in late-type 
disks \citep{Bigiel08}. 
For example, {using dust-based H$_2$ masses at spatial scales $\geq$\,1~kpc of 
the Small {Magellanic} Cloud}  \citet{Bolatto11} found $\tau_{\rm H2}$\,$\sim$~0.6--1.6~Gyr.
Also, \citet{Schruba12} found that {$\tau_{\rm H_2}$} for a sample of 16 nearby 
(D$\sim$\,4 Mpc) star-forming dwarf galaxies from the HERACLES {CO survey} 
\citep{Leroy08} is one or two orders of magnitude smaller than 
in normal spiral discs. 
They concluded that the inferred low values of $\tau_{\rm H2}$ may either 
indicate low H$_2$ masses coupled with high SFEs or that CO becomes a 
poor tracer of H$_2$ for these galaxies.  

In the above studies, however, most nearby dwarfs studied in 
detail are dwarf Irregulars (dI), while BCDs are strongly
underrepresented. 
Dwarf irregulars form stars at lower rates and their star formation 
histories are predominantly continuous, i.e., they are main sequence 
galaxies in the SFR-M$_{*}$ relation \citep[e.g.,][]{Brinchmann2004,Noeske2007,Lee2011,Hunt2012}. 
 {In contrast, BCDs show more extreme conditions \citep[e.g., high 
ISM ionization and higher {\it specific} SFR (sSFR);][]{Hunt2012,Amorin2014b,Amorin2015}, 
thus having predominantly bursty star formation histories \citep[e.g.,][]{MartinManjon12}. }
Therefore, one can argue that part of the enhancement in the SFE of
BCDs seen in the SK law is due to a more bursty SF history.  
{In this line, previous studies based on FIR/sub-mm data have 
found that even using a higher CO-H$2$ conversion factor, the depletion timescales 
for molecular gas in some metal-poor star forming galaxies appear considerably higher than 
for normal spirals \citep[e.g.,][]{Israel97,Gardan2007,Gratier2010a,Gratier2010b}. }
%

\begin{figure}[t!]
\centering
\includegraphics[width=8.5cm,angle=0]{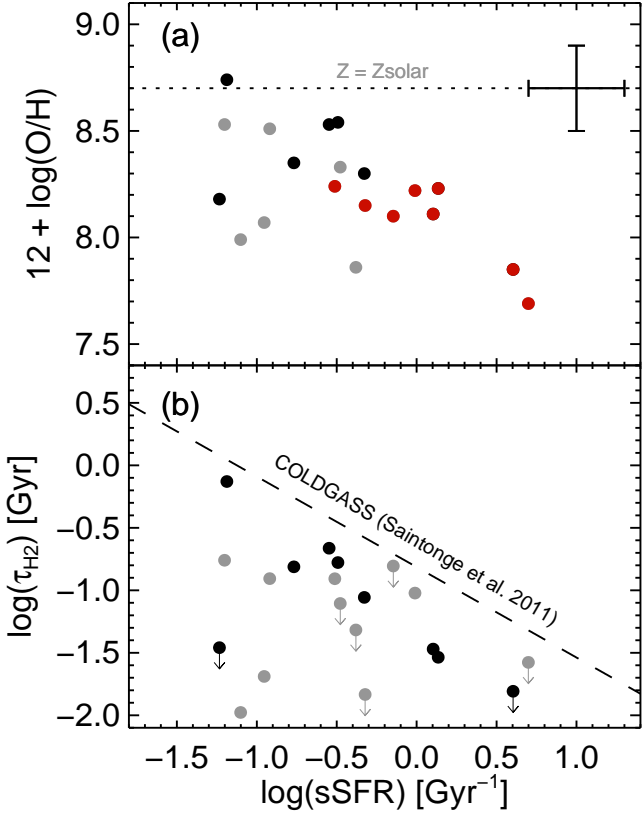}
\protect\caption[]{ {Metallicity ($a$) and H$_2$ depletion timescale ($b$)
  as a function of \textit{specific} SFR. Symbols are as in Fig.~\ref{fig8}, but 
red points in $a$ indicate BCDs with higher gas fractions 
($M_{\rm HI}/(M_{\rm HI}+M_{*}) > 0.5$) BCDs. The dotted line in $a$ indicates 
solar metallicity while the dashed line in $b$ shows the correlation found by 
\citet{Saintonge11b} for local star-forming disk galaxies of stellar masses 
M$_{*} \gtrsim$10$^{10}$\,M$_{\odot}$.}}    
\label{fig11}
\end{figure}

 {Following the above reasoning} in Fig.~\ref{fig11} we investigate the 
relation between sSFR and both H$_2$ depletion timescale and metallicity. 
On average,  {we find that} high sSFR BCDs  {have lower metallicity} 
(Fig.~\ref{fig11}$a$). These high sSFR galaxies  {also show} higher 
H{\sc i} gas mass fractions  {(red circles)}. 
Despite the large scatter, Fig.~\ref{fig11}$b$  {suggests} that BCDs 
with higher sSFR tend to have shorter $\tau_{\rm H_2}$.  
A similar trend  {was presented} by \citet[][dashed line 
in Fig.~\ref{fig11}$b$]{Saintonge11b}\footnote{The relation given by 
\citet{Saintonge11b} does not include Helium in $\Sigma_{\rm H_2}$,
so we have corrected upwards their relation to be consistent with our 
measurements.} for more massive (M$_{\star}\gtrsim$\,10 M$^{_\odot}$) 
star-forming galaxies of the COLDGASS survey \citep{Saintonge11a}. 
 {In their sample, however, the sSFR is not correlated with metallicity}. 
 {In Fig.~\ref{fig11}$b$ our BCDs follow the slope of the
relation for the COLDGASS sample}, but they  {appear shifted to}
significantly shorter H$_2$ depletion timescales for a given sSFR. 
 {These results} are consistent with \citet[][]{Leroy13}, who 
found that the eight less massive ($<$\,10$^{10}$\,M$_{\odot}$) galaxies in the 
HERACLES survey are systematically offset to larger $\tau_{\rm H_2}$ for a given 
sSFR when compared to more massive galaxies. 

 Taken together, results from Fig.~\ref{fig8} and Fig.~\ref{fig11} are 
mutually consistent and allow us to conclude that less massive and more 
metal-poor gas-rich galaxies, which form stars at higher rates, are those 
with shorter H$_2$ depletion timescales.

 {In the next section we use theses findings to derive a CO-to-H$_2$ 
conversion factor for BCDs, derive their H$_2$ masses, and revisit the scaling 
relations between star formation, gas content and metallicity.}  

\section{Deriving a metallicity-dependent CO-to-H$_2$ conversion
  factor for starbursting dwarfs}
\label{s7}

 {In order to derive an expression for a metallicity-dependent
  CO-to-H$_2$ conversion factor we use that $\alpha_{\rm
    CO, Z}/\alpha_{\rm CO, MW}$\,$\equiv$\,$\tau_{H_2}$\,$\times$\,$\Sigma_{\rm SFR}/\Sigma_{\rm H_2}$, 
    where $\tau_{\rm H_2}$ is the expected galaxy-averaged H$_2$ 
    depletion timescale for which metallicity effects have been accounted for. 
     Using the relation found between $\Sigma_{\rm SFR}/\Sigma_{\rm H_2}$ and metallicity 
  (Table~\ref{correlations}), we arrive to the following expression}, 

\begin{equation}\label{eq2}
\log (\alpha_{\rm CO, Z}/\alpha_{\rm CO, MW}) = \log{(2.4 \tau_{H_2})} - 1.91 (12+\log{\rm (O/H)} - 8.7) 
\end{equation}
 {with a 1$\sigma$ dispersion of 0.32 in the slope}.
Possible systematic uncertainties not included in Eq.~\ref{eq2}, e.g., those 
propagated from the derivation of integrated properties and driven by 
the numerous assumptions made, are likely larger and have not been taken into account. 

 {In order to derive a suitable value for the CO-to-H$_2$ conversion factor 
Eq.~\ref{eq2} yet requires the use of an appropriate value for $\tau_{\rm H_2}$. 
One possibility would be to derive H$_2$ masses from dust masses by assuming a dust-to-gas 
(D/G) ratio, and then compute $\tau_{\rm H_2}$. Especially in compact low-metallicity 
galaxies this approach is however not free of large uncertainties, which are mostly 
driven by variations in the D/G ratio with the ISM density \citep{Schruba12}, uncertainties 
in its calibration at different metallicities, and variations with the SF history of 
each galaxy \citep[][see also Bolatto et al. 2013]{Cormier2014,RemyRuyer2014}. 
In our case, only $\sim$\,20\% of the BCD sample have Herschel observations reported 
so far \citep[see][]{RemyRuyer2015}, thus precluding a complete analysis based on 
the dust masses.}

 {Another method recently used in the literature relies on the underlying 
assumption of a constant SFE ($\tau_{\rm H_2}$) for all galaxies 
\citep[e.g.,][see also McQuinn et al. 2012]{Schruba12,Genzel12}. This method has been 
motivated by the relatively constant SFE found observations of nearby disk galaxies 
\citep{Bigiel08,Bigiel11,Leroy08} and recent theoretical work \citep[e.g.,][]{Krumholz11} 
showing that the H$_2$ gas depletion timescale is uniform and does not depend on metallicity.   
However, in view of Fig.~\ref{fig11}, the hypothesis of a constant $\tau_{\rm H_2}$ 
appear not particularly appropriate for dwarf galaxies with predominately bursty SF histories.}

 {Therefore, in this work we choose an alternative method for obtaining $\tau_{\rm H_2}$ 
for each galaxy assuming that this quantity should follow a well established linear 
relation with sSFR at all masses \citep[e.g.,][]{Saintonge11b,Boselli2014,Huang2014}. 
In particular, here we use the relation derived by \citet{Saintonge11b} for star forming 
galaxies of stellar masses M$\gtrsim$\,10$^{10}$\,M${_\odot}$ in the COLDGAS survey 
(Fig.~\ref{fig11}), under the reasonable assumption that it extends to lower stellar masses 
(dashed line in Fig.~\ref{fig11}). Thus, we obtain the expected $\tau_{\rm H_2}$ from the 
ratio between the observed $\tau_{\rm H_2}$ computed via the galactic CO-to-H$_2$ conversion 
factor and the linear fit of \citet{Saintonge11b} for a given sSFR (Fig.~\ref{fig11}\,$b$). 
Finally, we use this value and Eq.~\ref{eq2} and derive $\alpha_{\rm CO}(Z)$ for each galaxy.
}

\begin{figure*}[t!]
\centering
\includegraphics[width=14cm,angle=0]{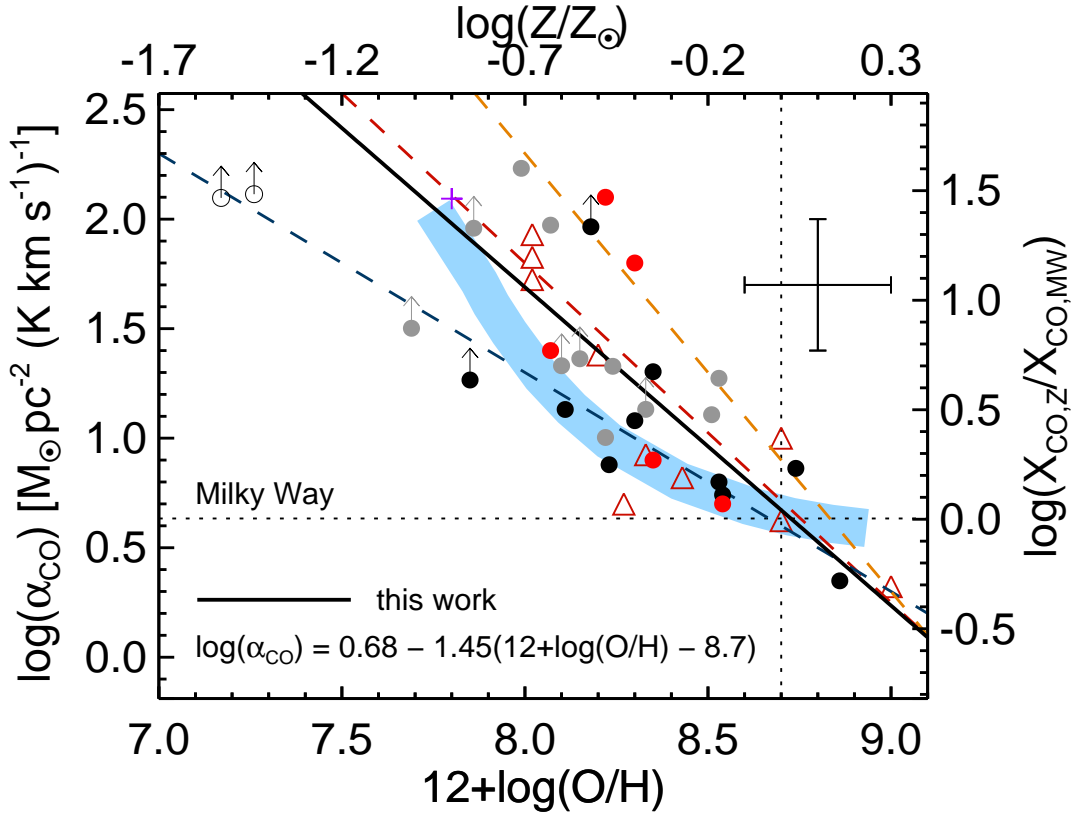}
\protect\caption[]{ {CO-to-H$_2$ conversion factor, $\alpha_{\rm CO}$, 
 as a function of the gas-phase metallicity. The right $y-$axis shows the ratio between 
 the metallicity-dependent conversion factor and the constant value
 for the Milky Way (horizontal dashed line). A dotted
 {vertical} line indicates {the} solar metallicity. 
Black and grey dots show \textit{subsample~I} and \textit{subsample~II} BCDs, 
respectively, while the black solid line shows their best linear fit, which is indicated 
in the legend. {Error bars indicate typical uncertainties for our sample.}
{Red dots indicate dust-based measurements for
  five BCDs in our sample with \textit{Herschel} data (see text)}. 
Triangles show dust-based measurements for Local Volume galaxies from 
\citet{Leroy11}, open circles are lower limits for the extremely metal-poor BCDs 
SBS\,0335-052 and I\,Zw\,18 \citep{Hunt2014}, and the magenta cross {corresponds} to the 
dIrr galaxy WLM from \citet{Elmegreen2013}.  
Blue, red, and orange dashed lines show best-fits for the same relation given by 
\citet{Arimoto96}, \citet{Genzel12}, and \citet{Schruba12}, respectively. 
The model  prediction of \citet{Wolfire10} (as presented in \citet{Sandstrom2013}) is shown 
{as a} light blue {band}. }
}    
\label{fig9}
\end{figure*}

 In Figure~\ref{fig9} we show the derived $\alpha_{\rm CO}$ as a function of 
metallicity. Despite the large scatter $\log \alpha_{\rm CO}$ appear to scale linearly 
with metallicity, meaning that $\alpha_{\rm CO}
\propto$\,($Z/Z_{\odot}$)$^{-1.5}$.  
{This relation is in qualitative agreement with previous 
  determinations, dust-based measurements, and recent model
  predictions, as we discuss below.  Fig.~\ref{fig9} suggests 
  that in vigorously star-forming dwarfs the fraction of H$_2$ traced
  by CO decreases a factor of about 40 from $Z \sim Z_{\odot}$ to $Z \sim 0.1 Z_{\odot}$, 
  leading to a strong underestimation of the H$_2$ mass in metal-poor systems when a 
  Galactic $\alpha_{\rm CO, MW}$ is considered}. 

 {In Figure~\ref{fig9} we have included dust-based
   $\alpha_{\rm CO}$ measurements for the five BCDs in the sample with
   Herschel data (red dots). We have used the dust masses compiled by
   \citet{RemyRuyer2014,RemyRuyer2015} to derive the H$_2$ masses following
   \citet{Cormier2014} and assuming the metallicity dependent D/G ratio of
   \citet{RemyRuyer2015}.
The resulting dust-based M$_{\rm H_2}$ and $\alpha_{\rm CO, dust}$ are
a factor of $\sim$\,1.5-30 higher than those derived using the
Galactic $\alpha_{\rm CO, MW}$. 
Within the errors (larger than a factor of 2), the dust-based measurements are
consistent with those based on our method and with the best-fit
relation ($\alpha_{\rm CO, Z}
\propto$\,($Z/Z_{\odot}$)$^{-1.5}$) shown in Figure~\ref{fig9}}. 

Similarly, our relation is broadly consistent with independent measurements of 
$\alpha_{\rm CO}$ based on different methods for {other} nearby star-forming galaxies of sub-solar 
metallicity. For example, our data are consistent with values based on dust modelling 
along the lines of sight from IR emission of five Local Group galaxies
(red triangles {in Fig.~\ref{fig9}}) by \citet{Leroy11}. 
Also, our power law index is $\sim$\,50\% higher than the one found by 
\citet{Arimoto96} based on the virial masses of giant molecular clouds of the Milky 
Way and eight additional nearby spirals and dIs (blue dashed line). 
Our index, instead, is in agreement within the uncertainties with an index of 
$\sim$\,1.6 derived by \citet{Genzel12} for more massive star-forming
galaxies at $z\leq$\,1 assuming that galaxies follow a universal SK
law of $n=$\,1.3 (red line {in Fig.~\ref{fig9}}). 
Also {\citet{Schruba12} studied the $\alpha_{\rm CO}-Z$ relation for
five HERACLES dwarf galaxies assuming a constant value for $\tau_{\rm
  H_2}=$\,1.8 Gyr. They found a power law index $\gtrsim$\,2
(orange line in Fig.~\ref{fig9}), significantly higher than ours}.  

 {Although the above results are qualitatively similar, quantitative differences 
are likely associated {with} the underlying assumptions behind each method, biases towards 
a certain galaxy class, and different dynamic ranges in metallicity, and uncertainties 
in its determination. In particular, the method based on a constant $\tau_{\rm H_2}$ has 
the caveat that it might mask-out any true variation of the SFE with {sSFR} at different 
metallicities. }

  {Finally, we find that in spite {of} the different parametrization 
 (power-law vs. exponential) our $\alpha_{\rm CO}-Z$ relation is broadly consistent 
 with model predictions\footnote{We use models from \citet{Wolfire10} normalized to 
 $\alpha_{\rm CO}=$\,4.4\,M$_{\odot}$\,pc$^{-2}$\,(K km s$^{-1}$)$^{-1}$ at solar 
 metallicity, as presented in \citet[][see also Leroy et al. (2013)]{Sandstrom2013}. 
 The models assume a fixed gas mass surface density for molecular clouds and a linear 
 scaling between the dust-to-gas ratio and metallicity.} of \citet{Wolfire10} from solar 
 to $\sim$\,10\% solar metallicities. }

\subsection{Revisiting the molecular content and star formation laws}
\label{s7.1}

\begin{figure*}[ht!]
\centering
\includegraphics[width=15cm,angle=0]{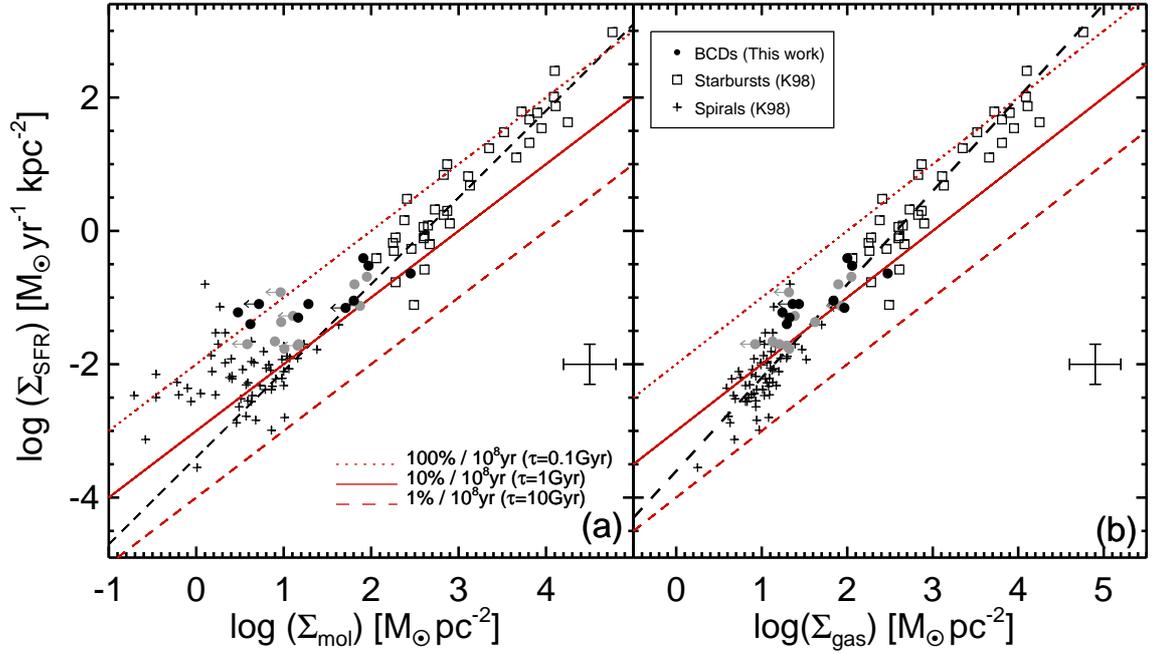}
\protect\caption[Corrected SK law]{ {Star formation rate surface density as a function 
of the expected H$_2$ (\textit{Left}) and H{\sc i}$+$H$_2$ (\textit{Right}) gas surface 
densities {after adopting the derived metallicity-dependent CO-to-H$_2$
  conversion factor $\alpha_{\rm CO, Z}$ from Eq.~2
  (Fig.~\ref{fig9})}. {Lines and symbols are as in Fig.~\ref{fig7}.}
}}
\label{fig12}
\end{figure*}

 {We use now the metallicity dependent CO-to-H$_2$ conversion factor
   $\alpha_{\rm CO, Z}$ obtained in the previous section in order to
   recompute H$_2$ masses and surface densities\footnote{In order to distinguish these 
quantities from those derived through a Galactic CO-to-H$_2$ conversion factor, we have adopted a 
different subscript (\textit{mol}), so M$_{\rm mol}$ and $\Sigma_{\rm mol}$.}. 
Using these quantities, we revisit different scaling relations presented before  
(Figs.~\ref{fig7}-\ref{fig11}). 
Thus, in Fig.~\ref{fig12} we show the corrected position of BCDs in the SK-law, while   
in Fig.~\ref{fig14} we show the same quantities as in Fig.~\ref{fig8} but using 
$\alpha_{\rm CO, Z}$.  In Figs.~\ref{fig14}\,$c-d$ we have included the 
predictions of models by \citet[][magenta and blue lines, respectively]{Krumholz11}.}

\begin{figure}[t!]
\centering
\includegraphics[width=8cm,angle=0]{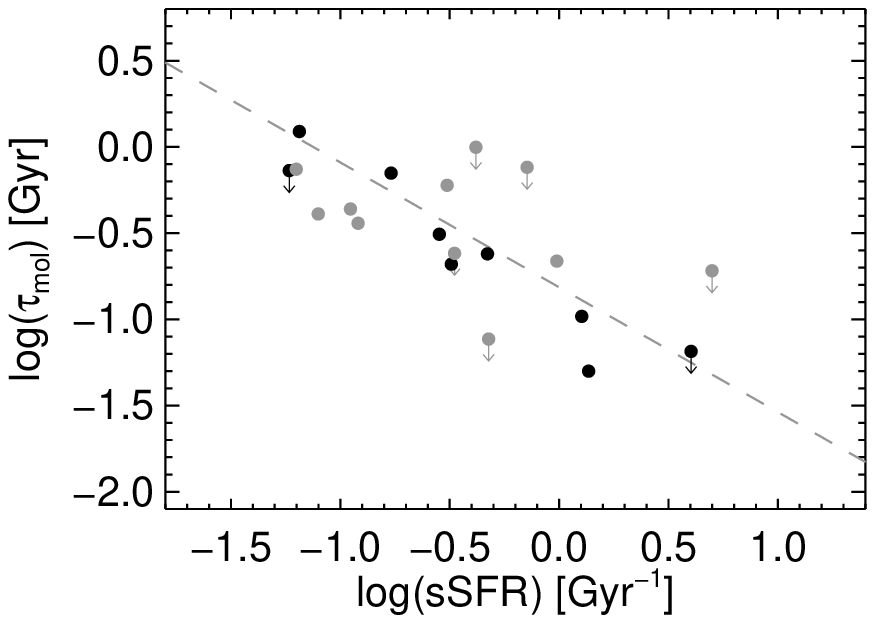}
\protect\caption[Corrected SSFR_tau]{Molecular depletion timescale, as derived 
using $\alpha_{\rm CO,Z}$ from Eq.~\ref{eq2} (see text), as a function of the 
\textit{specific} SFR. Symbols and line are as in Fig.~\ref{fig11}$b$.  }    
\label{fig13}
\end{figure}

%
 {We find that the use of a metallicity-dependent CO-to-H$_2$ conversion factor 
$\alpha_{\rm CO}(Z)$ from Eq.~\ref{eq2} {removes} only part} of
the large offset position observed 
for BCDs in  the SK law  {for molecular and total (H{\sc i}$+$H$_2$) gas followed by more 
massive starbursts and late-type galaxies. In particular, our results suggest 
that BCDs have shorter depletion timescales compared to normal late-type disks.
Starbursting dwarfs with larger gas fractions, low metallicity and higher SSFRs still show 
shorter depletion timescales for the molecular phase. This conclusion arises from 
Fig.~\ref{fig13} and Fig.~\ref{fig14}\,$c$. 
While some galaxies seem to agree with 
\citet{Krumholz11} models (magenta lines) predicting a constant 
SFE(H$_2$) with metallicity, some metal-poor dwarfs show a significantly increased efficiency. 
We note, however, that this result is in qualitative agreement with other model predictions 
\citep{P&P09,Dib11} which predict pronounced variations of $\tau_{\rm H_2}$ with metallicity 
during brief periods of intense star formation. 
BCDs with shorter molecular depletion timescales tend to show lower molecular fractions and higher 
sSFR, as shown in Fig.~\ref{fig14}\,$a-b$ and Fig.~\ref{fig13}. The depletion timescales 
for molecular plus atomic gas, instead, do not appear to be a strong function of metallicity, 
in qualitative agreement with the models of \citep{Krumholz11} (blue lines).
 }

%

\section{Summary and conclusions}
\label{s8}

We have studied the molecular content of a large and representative sample 
of {21} Blue Compact Dwarf galaxies selected from a  series of previous works. 
To this end, we have conducted new CO (1-0) and (2-1) single-dish observations  
of a sub-sample of 10 BCDs using the IRAM-30m telescope, further supplemented 
with similar data from the literature for the remaining {11} BCDs. 
Our CO observations have yielded 7 ($>$\,3$\sigma$) detections, one marginal 
($\sim$\,3$\sigma$) detection in CO(2-1), and two non-detections. 
For two BCDs (III\,Zw\,107 and Mrk\,401) CO  {emission has been detected} 
for the first time. 
The derived CO luminosities, in combination with an extensive ancillary data set, 
 {have been used} to study their relation to several galaxy-averaged properties, 
including SFR tracers, stellar and H{\sc i} masses, sizes and metallicity, 
as well as their impact on the star formation laws. 
We summarize our conclusions in the following:
\begin{figure*}[ht!]
\centering
\includegraphics[width=15.cm,angle=0]{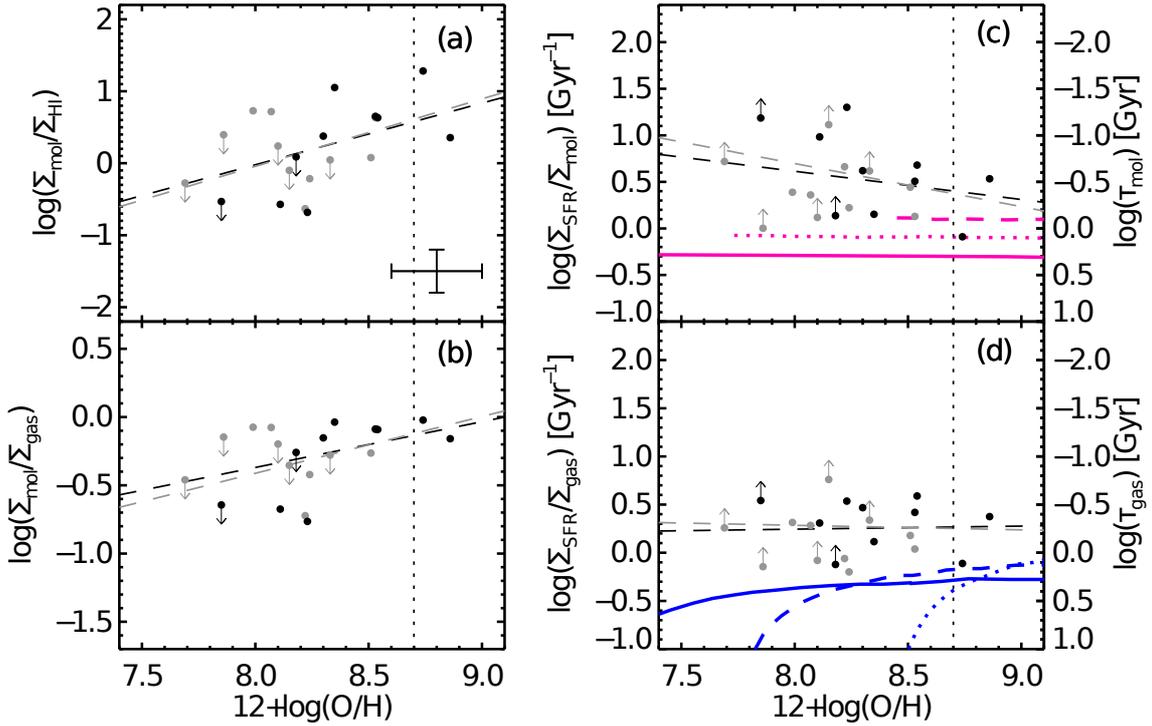}
\protect\caption[]{Same as Fig.~\ref{fig8}, but using the molecular masses derived 
using $\alpha_{\rm CO}(Z)$ from Eq.~\ref{eq2} (see text).  
Black and grey dashed lines show linear best-fits to {\it subsample~I} and 
{\it subsample~II}. Solid, dotted and dashed magenta (blue) lines in panel $c$ ($d$) 
show model predictions by \citet{Krumholz11} for different values of $\Sigma_{\rm gas}$ 
(80, 20 and 5  M$_{\odot}$\,pc$^{-2}$, respectively).
}    
\label{fig14}
\end{figure*}

\begin{itemize}
\item 
The amount of molecular mass traced by CO in BCDs scales with both stellar and 
H{\sc i} gas masses and all SFR tracers studied so far, in good agreement with 
previous findings for dwarf galaxies in general. 
Our results suggest that, over galaxy-wide scales, more 
massive and luminous BCDs -- which are those with strong and extended  
star-forming regions -- are favoured for CO detections. 
However, CO luminosity is a strong function of the gas-phase metallicity, 
which puts the most stringent constraint ($Z \ga$\,0.3\,$Z_{\odot}$) to 
the CO detectability in BCDs. 

\item 
In the context of the Schmidt-Kennicutt law, BCDs show larger SFR surface 
densities compared with other late-type galaxies. Thus, they appear 
systematically offset to lower H$_2$ and H$_{2}+$H{\sc i} depletion timescales 
from the general trend followed by dwarfs irregulars, spiral discs and more 
massive and dusty starbursts. 
We find a strong correlation between the BCD metallicity and both the
molecular gas fraction and the molecular gas depletion timescale; 
more metal-poor BCDs show lower molecular gas fractions and shorter 
depletion timescales than more metal-rich BCDs. 
However, for the \textit{total} (molecular plus atomic) gas depletion 
timescale the metallicity dependence becomes significantly reduced.  
Overall, we find that more metal-poor, gas-rich, and high sSFR BCDs are 
those with shorter H$_2$ depletion timescales, under the assumption of a 
constant (Milky Way) CO-to-H$_2$ conversion factor.  

\item
We discuss the above results in the context of the CO-to-H$_2$
conversion factor $\alpha_{\rm CO}$. 
{We have derived a metallicity-dependent CO-to-H$_2$
     conversion factor using} the relation between $\tau_{\rm H_2}$ and metallicity,  
{under the assumption} that the empirical correlation between the specific SFR and 
$\tau_{\rm H_2}$, found more metal-rich galaxies in the COLDMASS survey \citep{Saintonge11b} 
{can be extended} to lower masses. The result is
{$\log(\alpha_{\rm CO, Z})=0.7-1.5\times(12+\log(O/H)-8.7)$ 
(or $\alpha_{\rm CO, Z} \propto (Z/Z_{\odot})^{-y}$, with $y=1.5\,\pm\,0.3$}). 
  This power law is in qualitative agreement with previous 
  determinations based on dust-based H$_2$ masses and the model predictions 
  of \citet{Wolfire10}. Our power law index is however higher than values 
  based on virial masses of molecular clouds, and slightly lower than values 
  found assuming an universal $\tau_{\rm H_2}$ \citep[e.g.,][]{Schruba12,Genzel12}.
This result suggest that in vigorously star-forming dwarfs the 
fraction of H$_2$ traced by CO decreases a factor of about 40 from 
$Z \sim Z_{\odot}$ to $Z \sim 0.1 Z_{\odot}$, leading to a strong 
underestimation of the H$_2$ mass in metal-poor systems when a 
Galactic $X_{\rm CO, MW}$ is considered. 
 This supports previous work suggesting that in metal-poor and highly ionized environments 
 --such as those in high sSFR BCDs-- massive star formation is found 
 in increasingly CO-free molecular clouds. According to models 
 \citep[e.g.,][]{Wolfire10} 
 this is likely due to a rapid decrease of dust shielding and, consequently, 
 strong UV photodissociation \citep[see also][]{Bolatto13}.

\item
The observed relations between H$_2$, SFR and metallicity 
 {change when adopting the metallicity-dependent $\alpha_{\rm CO}$,   
alleviating the offset position of BCDs in the SK laws but not removing 
it completely. Our analysis suggest that \textit{starbursting dwarfs have shorter 
depletion gas timescales compared to normal late-type discs even accounting for the molecular gas not traced by CO emission in metal-poor environments}. 
Thus, $\tau_{\rm mol}$ {appears} not constant with metallicity but showing a small increase with 
metallicity mainly driven by galaxies with higher {sSFR} and gas fraction. 
While this produces some tension with some models \citep{Krumholz11} it 
{appears} to agree qualitatively with model predictions for brief and intense 
episodes of star formation \citep{P&P09,Dib11b}.  
While the revised relations between the molecular to atomic gas surface
density ratio and molecular fraction with {metallicity} show a shallower slope, 
the total (H$_2+$H{\sc i}) gas depletion timescale tend to become nearly constant 
with metallicity, in qualitative agreement with model 
predictions by \citet{Krumholz11}.}

\end{itemize}

 {We note that in the process of refereeing of the present paper a 
work by \citet{Hunt2015} tackling similar goals has just appeared published. 
Their results on the CO emission of 8 low-metallcity dwarf galaxies are certainly 
complementary to the work presented here, {and} their conclusions on the molecular 
depletion timescales of such galaxies appear in good agreement with ours.}

Further insights on the above conclusions would benefit from future 
studies using larger samples {of} star-forming dwarfs for which large and 
homogeneous multiwavelength datasets are currently available 
\citep[e.g., the AVOCADO project,][]{Sanchez-Janssen2013}. 

Being rare in the Local Universe, BCDs are one of the best local 
analogues of low-mass star-forming galaxies at high redshift. 
The strong limitations found {on} the use of CO as a suitable tracer of the 
molecular gas involved in the onset of starburst activity in nearby
BCDs clearly underline the strong challenge that will constitute to 
trace the molecular content of chemically unevolved galaxies in the
early Universe, even using the unprecedented sensitivity of ALMA.

\begin{acknowledgements}
 {The authors would like to thank the referee, J. Braine, for his 
helpful reports which significantly contributed to improving this manuscript.}
We are grateful to the staff of the IRAM 30m telescope for their support 
during the observations, and in particular to Sergio Martin. 
We thank U. Lisenfeld, J.M. V\'ilchez, P. Papaderos and S. Dib for helpful 
comments on the analysis. 
A significant part of this work has been presented in the 
PhD Thesis of R.~A. (Universidad de La Laguna, Spain, October 2008).
This work has been partially funded by the Spanish DGCyT, grants 
AYA2010-21887-C04-01, AYA2010-21887-C04-04, AYA2013-47742-C4-2-P,  
AYA2012-32295, AYA2013-43188-P, and FIS2012-32096.
This research was partially funded by the Spanish MEC under the 
Consolider-Ingenio 2010 Program grant CSD2006-00070: First Science 
with the GTC (http://www.iac.es/consolider-ingenio-gtc/).
R.~A. also {acknowledges} the contribution of the FP7 SPACE project “ASTRODEEP”
(Ref.No: 312725), supported by the European Commission. \\
This research has made use of the NASA/IPAC Extragalactic Database (NED) 
which is operated by the Jet Propulsion Laboratory, California Institute 
of Technology, under contract with the National Aeronautics and Space 
Administration.
 
\end{acknowledgements}

\end{document}